# Protein sequence classification using natural language processing techniques

Huma Perveen[1], Julie Weeds[2]

[1] School of Mathematical and Physical Sciences, University of Sussex, Brighton, UK

[2] School of Engineering and Informatics, University of Sussex, Brighton, UK

Corresponding author
E-mail: humaperveen@ymail.com

ORCiD: 0009-0007-6854-2155

The principal author is Huma Perveen

# Abstract

**Purpose:** This study aimed to enhance protein sequence classification using natural language processing (NLP) techniques while addressing the impact of sequence similarity on model performance. We compared various machine learning and deep learning models under two different data-splitting strategies: random splitting and ECOD family-based splitting, which ensures evolutionary-related sequences are grouped together.
**Methods:** The study evaluated models such as K-Nearest Neighbors (KNN), Multinomial Naïve Bayes, Logistic Regression, Multi-Layer Perceptron (MLP), Decision Tree, Random Forest, XGBoost, Voting and Stacking classifiers, Convolutional Neural Network (CNN), Long Short-Term Memory (LSTM), and transformer models (BertForSequenceClassification, DistilBERT, and ProtBert). Performance was tested using different amino acid ranges and sequence lengths with a focus on generalization across unseen evolutionary families.
**Results:** The Voting classifier achieved the highest performance with 74% accuracy, 74% weighted F1 score, and 65% macro F1 score under random splitting, while ProtBERT obtained 77% accuracy, 76% weighted F1 score, and 61% macro F1 score among transformer models. However, performance declined across all models when tested using ECOD-based splitting, revealing the impact of sequence similarity on classification performance.
**Conclusion:** Advanced NLP techniques, particularly ensemble methods like Voting classifiers, and transformer models show significant potential in protein classification, with sufficient training data and sequence similarity management being crucial for optimal performance. However, the use of biologically meaningful splitting methods, such as ECOD family-based splitting, is crucial for realistic performance evaluation and generalization to unseen evolutionary families.

# 1. Introduction

Proteins play crucial roles in living organisms, including catalyzing metabolic processes, replicating DNA, reacting to stimuli, providing structure, and transporting molecules. Proteins are composed of long chains of amino acids, and the sequence of these amino acids determines the protein's structure and function. Understanding the relationship between amino acid sequence and protein function has significant scientific implications, such as identifying errors in biological processes and clarifying protein synthesis mechanisms.

Traditional methods for determining protein functions, such as crystallography and biochemical studies, are time-consuming [1]. To enhance protein classification, we propose a system that utilizes machine learning, deep learning, and NLP techniques. NLP has emerged as a powerful tool for classifying protein sequences. By treating protein sequences like text, NLP techniques, such as n-grams and word embeddings (are a representation of text where words that have a similar meaning have a corresponding similar representation. Individual words are represented as real-valued vectors in a predefined vector space. The vector can be of tens or hundreds of dimensions), offer a unique approach to protein sequence analysis by treating amino acid sequences similarly to linguistic data, where each amino acid acts as a 'word.' This perspective allows us to capture sequence patterns effectively, enable efficient classification and functional prediction. Models like BERT enhance these processes by learning contextual relationships within the sequences, significantly improving the extraction of valuable insights from complex biological data. This study aims to assess the effectiveness of NLP techniques in classifying protein sequences into functional categories. We hypothesize that ensemble models and transformer-based approaches will outperform traditional models, especially on large and diverse datasets.

## 2. Related work

The study of proteins and other molecules to ascertain the function of many novel proteins has become the foundation of contemporary biological information science. Various methods have been developed to encode biological sequences into feature vectors and classify them using machine learning algorithms. In an experiment, Dongardive et al found that biological sequences are encoded into feature vectors using the N-gram algorithm. 717 sequences divided unevenly into seven classes make up the dataset used for the studies. The closest neighbors are determined using the Euclidean distance and the cosine coefficient similarity metrics [1]. In their 2017 study, Li M. et al. classified the protein sequences of GCPRs (G-protein Coupled Receptors). The dataset included 1019 different protein sequences from the GCPR superfamily. These sequences have all been examined in UniProtKB. The data was pre-processed, and the feature selection methods Term Frequency - Inverse Document Frequency (TF-IDF) and N-gram were utilized [2].

In their work, Lee T. and Nguyen T. used the analysis of unprocessed protein sequences to learn dense vector representation. The information was gathered from the 3,17,460 protein sequences and 589 families in the Universal Protein Resource (UniProt) database. Using Global Vectors for Word Representation (GloVe), a distributed representation was made by encoding each sequence as a collection of trigrams that overlapped [3]. According to Vazhayil A. et al., a protein family is a group of proteins that have the same functions and share similar structures at the molecular and sequence levels. Although a sizable number of sequences are known, it is noted that little is known about the functional characteristics of the protein sequences. Swiss-Prot's Protein Family Database (Pfam), which has 40433 protein sequences from 30 distinct families, was used as the source of the data for this study. Redundancy in the dataset was checked, and it was discovered that there were no redundant protein sequences. To represent discrete letters as vectors of continuous numbers, the text data was first processed using Keras word embedding and N-gram [4]. Islam et al. developed an NLP-based protein classification method that automates the feature generation process using a modified combination of n-grams and skip-grams (m-NGSG) [5].

Given the extensive applications in clinical proteomics and protein bioinformatics, protein sequence analysis has been a focus of numerous studies in recent years, including works by Barve [6], Chen [7], Cong [8], Machado [9], Carregari [10], and Liu [11]. This analysis is crucial for characterizing protein sequences and predicting their structures and functions. Comparative analysis of protein sequences has proven more sensitive than direct DNA comparison, leading to the establishment of multiple protein sequence databases, such as PIR, PDB, and UniProt.

Protein sequence classification, as highlighted by Baldi and Brunak, plays a pivotal role in these analyses because members of the same protein superfamily are often evolutionarily related and share functional and structural similarities. Correctly classifying a protein sequence into its superfamily streamlines molecular analysis within that group, reducing the need for exhaustive individual protein sequence analysis. Typically, two protein sequences are classified together if their features, identified through sequence alignment algorithms, exhibit high homology [12].

Various alignment algorithms have been developed to determine the class of an unknown protein sequence by comparing it with known sequences, calculating similarities, and using databases such as iPro-Class, SAM, and MEME. However, these comparisons can be time-consuming, especially with large databases or long sequences, making it crucial to develop efficient classification systems.

In recent decades, several methods have been employed for general signal classification based on statistical theory, such as decision trees, support vector machines (SVM), and neural networks (NN). Yang et al. utilized word segmentation for feature extraction and SVM for classification [13], while Caragea et al. employed hashing to reduce the dimensionality of protein sequence feature vectors before classification with SVM [14].

Neural networks have also become popular for protein sequence classification due to their ability to handle the high-dimensional, complex features of protein sequences. For example, Wang et al. proposed a modular radial basis function (RBF) neural network for protein sequences [15,16], while Wang and Huang applied a computationally efficient extreme learning machine (ELM) for single-layer feedforward neural networks (SLFNs) to this task [17-20]. Their results indicated that ELM significantly outperformed traditional gradient-based methods which was developed by Levenberg [21] and Marquardt [22] in both speed and classification rate.

To further enhance classification performance while maintaining reasonable training times, Cao et al. introduced a self-adaptive evolutionary ELM (SaE-ELM), which utilizes differential evolution to optimize the hidden neuron parameters in ELM networks [23].

Cao and Xiong investigated the protein sequence classification problem using SLFNs on the protein sequence datasets from the Protein Information Resource center (PIR) and introduced ensemble-based methods to enhance classification performance. Their study proposed two novel algorithms, voting based ELM (V-ELM) and voting based optimal pruned ELM (VOP-ELM), which outperformed existing state-of-the-art methods in accuracy. Particularly, VOP-ELM demonstrated the highest recognition rate, making it a significant advancement in protein sequence classification [24].

A few recent studies have pretrained deep neural language models on protein sequences e.g. Evolutionary Scale Modeling (ESM) [25], Tasks Assessing Protein Embeddings (TAPE-Transformer) [26], ProtTrans trained two auto-regressive models (Transformer-XL, XLNet) and four auto-encoder models (BERT, Albert, Electra, T5) on data from UniRef and BFD containing up to 393 billion amino acids [27]. By training both auto-regressive and auto-encoder models on massive protein datasets, recent studies have demonstrated that these protein-specific LMs (pLMs) can capture biophysical properties and make highly accurate predictions in tasks such as protein secondary structure and sub-cellular location, often outperforming traditional methods. This progress highlights the potential of pLMs in advancing protein sequence analysis [27].

Shinde et al considered the structural protein sequences with 10 classes and were able to get 90 % accuracy using a convolutional neural network [28]. A ResNet-based protein neural network architecture is proposed by Maxwell et al. for Pfam dataset families [29]. ProteinBERT is a deep language model created exclusively for proteins, according to Brandes et al. ProteinBERT's architecture combines local and global representations, enabling the processing of inputs and outputs from beginning to end. Even with limited labelled data, ProteinBERT offers an effective framework for quickly developing protein predictors [30]. The ProtBert model was pre-trained on Uniref100 [31] a dataset consisting of 217 million protein sequences [32]. Wang et al. proposed model effectively employs hierarchical attention mechanism and capsule networks for biomedical document triage task [33].

Recent studies showcase the transformative role of computational models in protein function prediction. For example, Ens-Deep-AGP uses advanced feature engineering and deep learning to achieve high accuracy in predicting angiogenic proteins, which are critical for therapies in cancer and heart disease [34]. Similarly, the VEGF-ERCNN model leverages GAN and ERCNN techniques to predict Vascular Endothelial Growth Factor (VEGF) with high precision, impacting vascular treatment strategies [35]. Epigenetic protein prediction models using ERCNN and encoding methods like DDE and GAAC also demonstrate strong predictive performance, accelerating drug discovery for autoimmune and neurological diseases [36]. Additionally, CL-Pred achieves high accuracy in identifying clathrin proteins, vital for endocytosis and signal transduction, suggesting its utility in drug target identification [37]. Some recent studies highlight computational models for identifying critical biological targets. Drug-LXGB, a predictor for druggable proteins (DPs), enhances DP identification using features from amino acid compositions and advanced descriptors. With extreme gradient boosting and 10-fold cross-validation, Drug-LXGB outperforms existing predictors, advancing drug discovery [38]. Additionally, BLP-piwiRNA improves piwiRNA prediction, a non-coding RNA with roles in gene regulation and cellular defense. Leveraging deep neural networks and optimal feature selection, it offers higher prediction accuracy, potentially aiding tumor diagnostics [39]. Lastly, Deep-AGP identifies angiogenic proteins (AGPs), crucial for cancer and cardiovascular treatments, using novel descriptors and 2D-CNNs [40]. These models are poised to impact therapeutic development significantly. Seo S. et al. introduced DeepFam, an alignment-free method that can extract functional information directly from sequences without the need of multiple sequence alignments [46].

Despite advancements in protein classification methods, existing techniques are often limited in scalability, precision, and handling of complex datasets with imbalanced classes. This study addresses these gaps by employing NLP to enable automated, scalable classification across 75 diverse protein classes. This study uses a more comprehensive dataset with 75 target protein classes from the PDB dataset, offering a more extensive exploration of protein sequence classification compare to [1,2]. Unlike the single approach used by Dongardive (N-gram algorithm) and Shinde (CNN), this work explores a wide range of machine learning and deep learning models, including KNN, Naive Bayes, Logistic Regression, Decision Trees, Random Forest, XGBoost, CNN, LSTM and advanced transformers like ProtBert, providing a broader evaluation of techniques. This study is unique in that it employs ensemble methods (Voting and Stacking classifiers), which were not explored in the mentioned studies. These methods achieved superior performance, showing the value of combining models to improve classification accuracy and robustness. While Shinde et al. used CNN and Dongardive et al.

focused on N-grams, this work integrates state-of-the-art transformer models (BertForSequenceClassification, DistilBERT, and ProtBert), which are pre-trained on vast protein sequence datasets, allowing for better contextual understanding and performance in sequence classification. This study also explores a range of amino acid N-grams (1-4 grams) for machine learning models, and emphasises the importance of sequence similarity across different classes, which is a critical issue in protein classification, and how AI models can handle these similarities. To prevent data leakage, we utilized Evolutionary Classification of Protein Domains (ECOD) [47,48], a hierarchical classification system that organizes protein domains based on their evolutionary relationships. Unlike structure-based classifications such as SCOP and CATH, ECOD focuses on detecting remote homology between protein domains and is continuously updated to classify newly released structures in the PDB archive. By leveraging ECOD, we ensured that protein sequences from the same evolutionary family were grouped together, providing a biologically meaningful data split and preventing information leakage between training and test sets. This challenge is not directly addressed in the cited studies. Table 1 presents a comparison of various existing research in protein sequence classification.

**Table 1 Comparison of related studies for protein sequence classification**

| Dataset | Applied Techniques | Best Result | Authors and References |
|---|---|---|---|
| 717 protein sequences | K-Nearest Neighbor (KNN) | KNN with 84% | Dongardive et al (2016) [1] |
| G-protein Coupled Receptors (GPCRs) super family | Convolutional Neural Network (CNN) | an accuracy of up to 98.34%, 98.13% and 96.47% in the classification of family level, subfamily level I and II, respectively | Li M. et al. (2017) [2] |
| Protein Data Bank | Decision Tree, Random Forest, Extra Trees | Random Forest with 92.7% | Parikh Y. et al. (2019) [42] |
| PROSITE database, SCOP AND GPCRs | Support Vector Machine, Decision Tree, Naive Bayes | Decision Tree with 77.3%, SVM with 92.6 % and 88.8% for GPCRs subfamily level I and level II | Yang Y. et al. (2007) [43] |
| | | | |
| SCOP | Random Forest, K star, Nearest Neighbour, Logistic Regression, Support Vector Machine, Naïve Bayes | Random Forest with 73.7% | Li J. et al. (2013) [44] |
| Structural Protein Sequences | Convolutional Neural Network (CNN) | Convolutional Neural Network with 90% | Shinde A. and D'Silva M. (2019) [28] |
| 4prot and CB513 CullPDB | Convolutional Neural Network (CNN) | Convolutional Neural Network with 90.93% | Jalal S. et al. (2019) [45] |
| Clusters of Orthologous Groups (COGs) and GPCRs dataset | Convolutional Neural Network (CNN) | 91.7% on COGs dataset and 97.2%, 86.8% and 81.2% for family, sub-family and sub-subfamily on GPCRs dataset | Seo S. et al. (2018) [46] |
| UniRef90 | BERT Transformer model | Secondary structure 74% | Brandes et al. (2022) [30] |
| UniRef, BFD dataset | auto-regressive models (Transformer-XL, XLNet) and auto-encoder models (BERT, Albert, Electra, T5) | Secondary structure 81-87%, protein sub-cellular location 81%, solubility 91% | Elnaggar A et al. (2022) [32] |
| Structural Protein Sequences | KNN, Naive Bayes, Logistic Regression, Decision Trees, Random Forest, XGBoost, voting and stacking classifier, CNN, LSTM and transformer models (BertForSequenceClassification, DistilBERT, and ProtBert) | ProtBERT 77%, Voting classifier 74% | This Study |

# 3. Material and methods

The study employs NLP techniques including n-gram (range 1-4) models, word embedding-based deep learning approach (explained more in section 3.4.3 and Fig.

5) and transformers. N-grams help capture amino acid patterns, while transformers such as BERT variants facilitate sequence understanding by leveraging contextual embeddings.

## 3.1 Dataset

The dataset used in this study is the structural protein sequences dataset from Kaggle [41], derived from the Protein Data Bank (PDB) of the Research Collaboratory for Structural Bioinformatics (RCSB). It contains over 400,000 protein structural sequences, organized into two files: `pdb_data_no_dups.csv` (protein metadata) and `data_seq.csv` (protein sequences).

## 3.2 Exploratory data analysis (EDA)

The initial preprocessing steps involved merging the two CSV files on the structure ID column, removing duplicates, dropping null values, and selecting data where the macromolecule type is protein.

### 3.2.1 Sequence length analysis

After analyzing the statistics in (Fig. 1a) and observing the boxplot in (Fig. 1b), it can be concluded that the mean sequence character count is approximately 280. This length is considered optimal for applying BERT models, as it provides a sufficient amount of information for deep learning algorithms to process effectively. For the purpose of this study, sequences with more than 30 amino acids in length were considered. This threshold ensures that the sequences are long enough to capture meaningful patterns and interactions. The sequence character count is a crucial parameter in deciding the appropriate sequence length for deep learning applications, impacting the performance and accuracy of the models employed.

**Fig. 1 Sequence character count**

| | seq_char_count |
|---|---|
| count | 156068.000000 |
| mean | 278.822648 |
| std | 201.573606 |
| min | 31.000000 |
| 25% | 148.000000 |
| 50% | 241.000000 |
| 75% | 351.000000 |
| max | 5037.000000 |

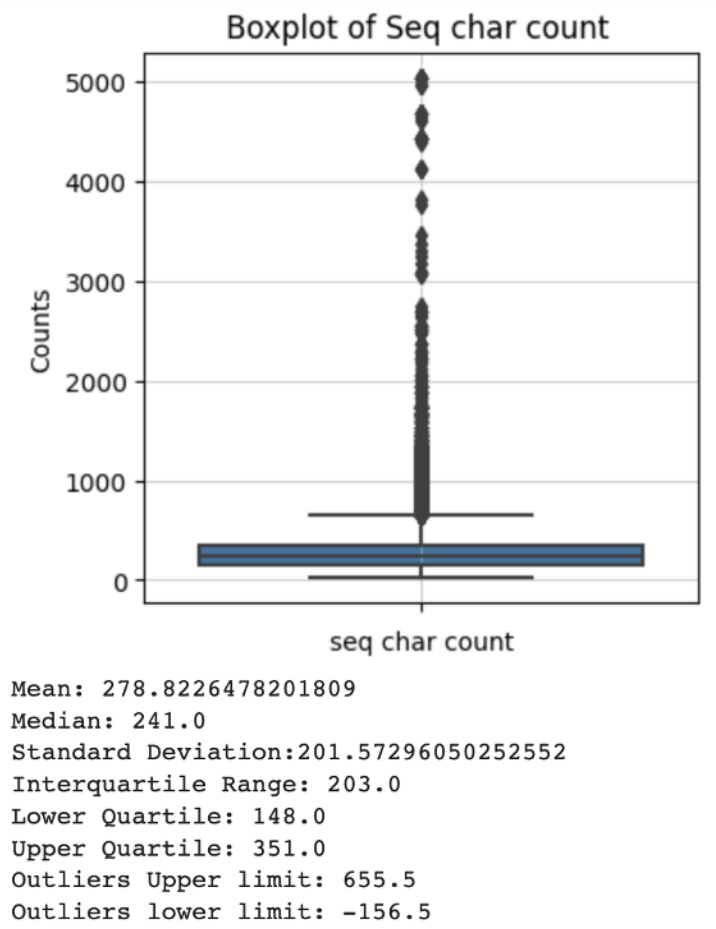

(a) Statistics of sequence character count (b) Boxplot of sequence character count

In our analysis of the sequence lengths, as depicted in (Fig. 2), it is evident that the distribution is highly skewed. The majority of unaligned amino acid sequences fall within a character count range of 50 to 450. This skewness indicates a significant variation in sequence lengths within the dataset, which could have implications for downstream analyses and the computational approaches employed. Understanding the distribution of sequence lengths is crucial for optimizing alignment algorithms and improving the accuracy of subsequent bioinformatics analyses.

**Fig. 2 Distribution of sequence length**

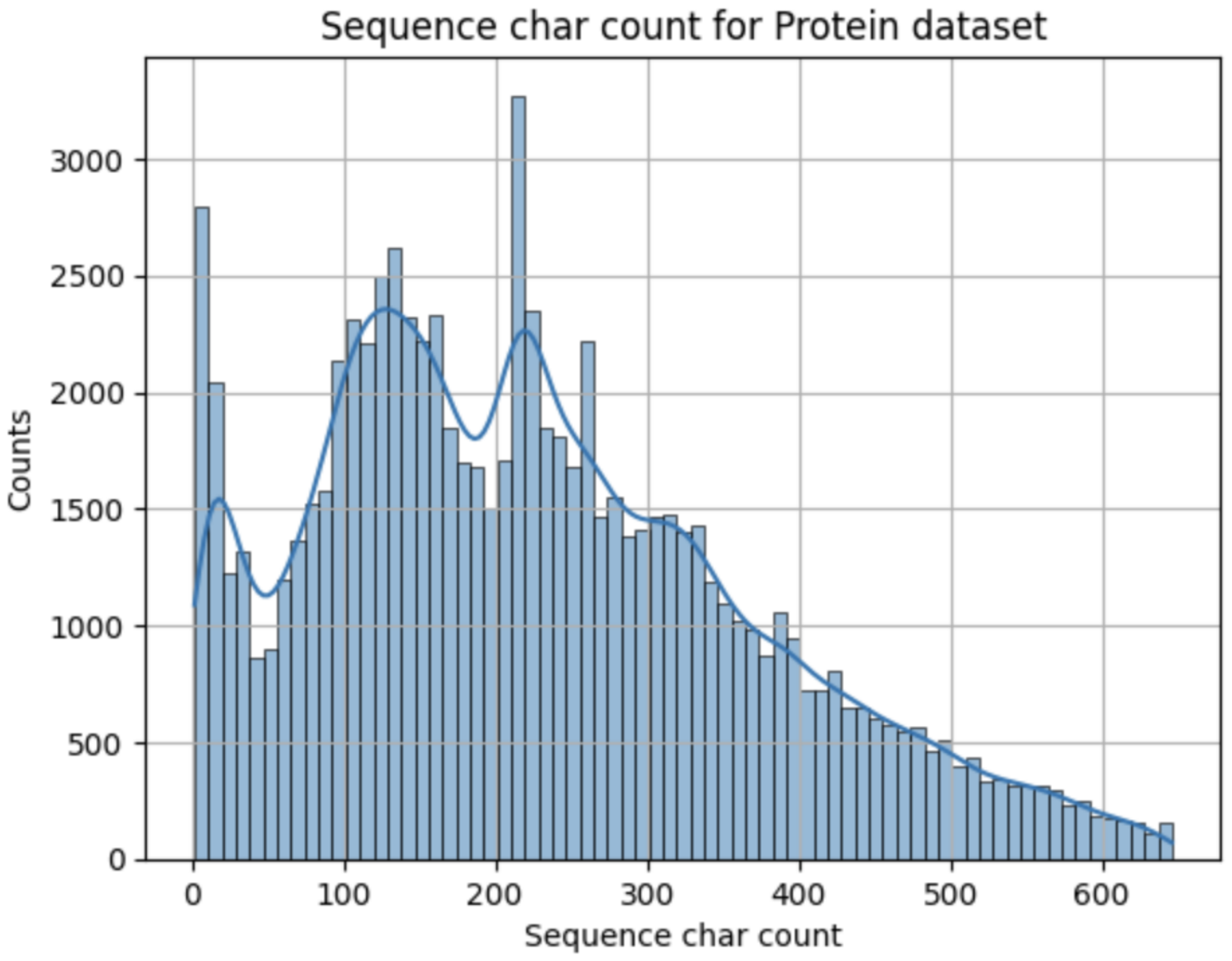

## 3.2.2 Amino acid frequency analysis

Amino acid sequences are represented with their corresponding 1-letter code, for example, the code for alanine is (A), arginine is (R), and so on.

In our analysis, Fig. 3 highlights the frequency distribution of amino acids in the dataset. It is evident that leucine (L) appears most frequently, succeeded by alanine (A), glycine (G), and valine (V). This observation aligns with the known biological abundance of these amino acids in various proteins. Additionally, our sequence encoding approach focused on the 20 standard amino acids, deliberately excluding the rare amino acids such as X (any amino acid), U (selenocysteine), B (asparagine or aspartic acid), O (pyrrolysine), and Z (glutamine or glutamic acid) to streamline the analysis and ensure consistency.

**Fig. 3 Amino acid distribution**

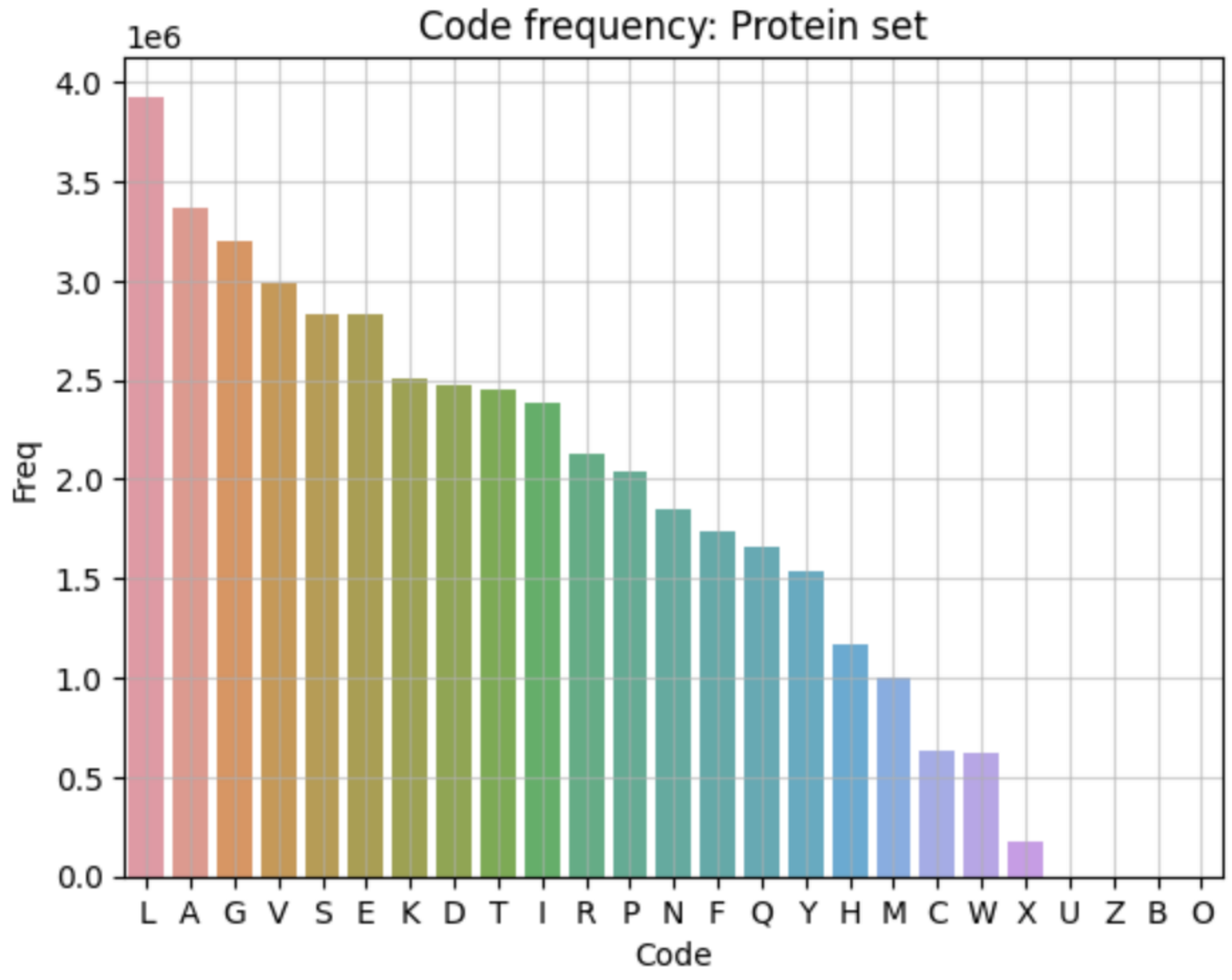

## 3.3 Machine learning methods

### 3.3.1 Data transformation

Sequences and classes were initially converted into numerical form using CountVectorizer and LabelEncoder. The dataset was then split into training and test sets with an 80:20 ratio. We compared various models under two different data-splitting strategies: **random splitting** and **ECOD family-based splitting**, which ensures evolutionary-related sequences are grouped together. To optimize the model's hyperparameters, RandomizedSearchCV with 3 fold cross validation was employed, ensuring the best possible configuration for each machine learning method, because it is more efficient if the search space is large since it only samples a subset of the possible combinations rather than evaluating them all.

### 3.3.2 Methods

Several machine learning algorithms were applied to the dataset. These included K-Nearest Neighbors (KNN), Multinomial Naive Bayes, Logistic Regression, Multilayer Perceptron, and Decision Tree Classifier. Additionally, various ensemble methods were utilized, such as Random Forest Classifier, XGBoost Classifier, Voting Classifier, and Stacking Classifier, to improve classification performance through model combination and aggregation. Estimator classifier applied for the voting algorithm are Logistic regression, K-Nearest Neighbor, Random Forest and XGBoost. Hyperparameter values for these classifiers are best values obtained during training individual model. StratifiedKFold is used to split the training set into 10 folds. Performance metrics are recorded for each fold. After cross-validation, the voting classifier is trained on the entire training set and evaluated on the separate test set. The results are better with soft voting strategy because it is the average probability of all the combined classifier. Estimator applied for the stacking algorithm are KNN, RFC and XGBoost, final estimator is XGBoost. Best hyperparameter values are mentioned in Table 2.

**Table 2 Hyperparameter for machine learning models**

| Models | Best Value |
|---|---|
| **3-gram KNN** | n_neighbors: 3, weights: 'distance', leaf_size: 13 |
| **4-gram Multinomial NB** | alpha: 1.0 |
| **4-gram LR** | Penalty: 'l2', C: '1.0', Solver: 'newton-cg', Class_weight: 'balanced' |
| **4-gram MLP** | Hidden_layer_sizes: (100,), alpha: '0.001', Learning_rate: 'invscaling' |
| **4-gram DTC** | criterion: 'gini', splitter: 'best', class_weight: 'balanced' |
| **4-gram RFC** | n_estimators: 100, criterion: 'gini', class_weight: 'balanced' |
| **2-gram XGB** | n_estimators: 500, max_depth: 7, Learning_rate: 0.1 |
| **4-gram Voting Soft** | Best value obtained for LR, KNN, RFC, XGB |
| **4-gram Stacking** | Best value obtained for KNN, RFC, XGB |

## 3.4 Deep learning methods

### 3.4.1 Further pre-processing of sequences with Keras

In our study, tokenization is a crucial step to prepare protein sequences for processing by deep learning models. Each amino acid in a sequence is converted into a unique integer, which transforms the biological sequence data into a numerical format that can be processed by neural networks. Further pre-processing of sequences was conducted using Keras. For instance, a sequence 'GSAFCNLARCELSCRSLGLLGKCIGEECKCVPY' will be converted into encoding like this [ 6, 16, 1, 5, 2, 12, 10, 1, 15, 2, 4, 10, 16, 2, 15, 16, 10, 6, 10, 10, 6, 9, 2, 8, 6, 4, 4, 2, 9, 2, 18, 13, 20]. This encoding enables models to identify and learn patterns within the protein sequences, similar to how words are processed in natural language processing (NLP).

To ensure uniform input length, each sequence is padded to pre-defined lengths of 100, 256, and 512, allowing us to evaluate model efficiency across varying sequence lengths. These lengths were chosen to capture different levels of context within the sequence:

- **100**: Shorter sequences to test for efficiency and computational speed.
- **256 and 512**: Longer sequences to preserve more biological information, allowing the models to capture distant dependencies within the amino acid sequence.

By experimenting with different lengths, we aimed to optimize between computational resources and model accuracy, balancing sequence length with model complexity.

The padded sequence would be like the below:

[ 6, 16, 1, 5, 2, 12, 10, 1, 15, 2, 4, 10, 16, 2, 15, 16, 10, 6, 10, 10, 6, 9, 2, 8, 6, 4, 4, 2, 9, 2, 18, 13, 20, 0, 0, 0, 0, 0, 0, 0, 0, 0, 0, 0, 0, 0, 0, 0, 0, 0, 0, 0, 0, 0, 0, 0, 0, 0, 0, 0, 0, 0, 0, 0, 0, 0, 0, 0, 0, 0, 0, 0, 0, 0, 0, 0, 0, 0, 0, 0, 0, 0, 0, 0, 0, 0, 0, 0, 0, 0, 0, 0, 0, 0, 0, 0, 0, 0, 0, 0]

where trailing zeros ensure consistent input size across samples.

To prevent data leakage, we experimented two different data-splitting strategies: **random splitting** (The data was split into training, validation, and test sets in a 70:10:20 ratio) and **ECOD family-based splitting**. For class transformation, sequences with counts greater than 100 were selected, and labels were transformed to one-hot representation using LabelBinarizer. Class weights were assigned using sklearn's compute_class_weight module to address class imbalance. Early stopping was implemented as a regularization technique to prevent overfitting, with performance monitored after each epoch. The deep learning models used included Long Short-Term Memory (LSTM) and Convolutional Neural Network 1D (CNN 1D).

### 3.4.2 Long short-term memory (LSTM)

LSTM (Long Short-Term Memory) is a type of recurrent neural network (RNN). that is designed to effectively capture and model long-term dependencies in sequential data. It is particularly useful in tasks involving sequential data, such as natural language processing, speech recognition, and time series analysis. The LSTM architecture introduces memory cells and gating mechanisms to overcome the limitations of traditional RNNs, which struggle with capturing long-term dependencies due to the vanishing gradient problem. The memory cell is the core component of an LSTM. It allows the network to store and access information over long sequences. The memory cell maintains an internal state, which is updated and modified as new input is processed. LSTMs use three types of gates to control the flow of information: the input gate, the forget gate, and the output gate. These gates are responsible for regulating the information flow into and out of the memory cell. The cell state is the internal memory of the LSTM. It runs through time and carries information from one time step to another, allowing the LSTM to capture long-term dependencies. The hidden state is the output of the LSTM at each time step. It is a transformed version of the cell state and carries information that is relevant for the current prediction or task. The hidden state is computed based on the cell state and the output gate.

- **Architecture and Embedding**: The LSTM model begins with an embedding layer that maps each tokenized amino acid to a dense vector, creating a more meaningful representation of each sequence. This helps the model capture relationships and dependencies between amino acids.
- **LSTM Cells and Temporal Dependencies**: LSTM cells are designed to capture long-range dependencies in sequential data. As each amino acid passes through the LSTM cells, the model retains context, allowing it to learn relationships over the sequence, which is essential for capturing functional patterns within proteins.
- **CuDNNLSTM for Efficiency**: To optimize training speed, we used CuDNNLSTM, a GPU-accelerated version of LSTM, which significantly reduces training time while maintaining model accuracy.
- **Dropout Regularization**: A dropout layer with a rate of 0.2 is applied to prevent overfitting by randomly "dropping" 20% of neurons during each training step. This enhances the model's generalizability.
- **Dense Output Layer**: Finally, a dense layer with softmax activation outputs the probabilities for each protein class, enabling multiclass classification.
- The model is trained using categorical cross entropy and is compiled using Adam optimizer.

An outline of proposed implementation steps is drawn in (Fig. 4).

**Fig. 4 LSTM implementation outline**

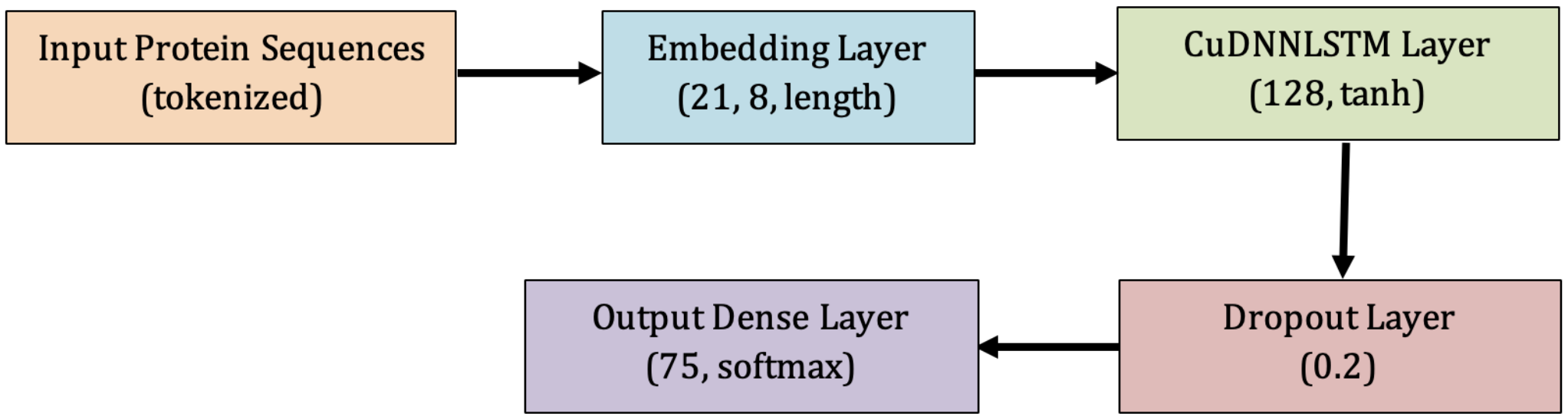

### 3.4.3 Convolutional neural network 1D (CNN 1D)

Recent success in NLP suggests using word embeddings which are already implemented as a Keras Embedding layer. Note that in this dataset, there are only 20 different words (for each amino acid). Instead of using every n-gram, using 1D-convolution on the embedded sequences is considered. The size of the convolutional kernel can be seen as the size of n-grams and the number of filters as the number of words as shown in (Fig. 5).

**Fig. 5 CNN1D mechanism**

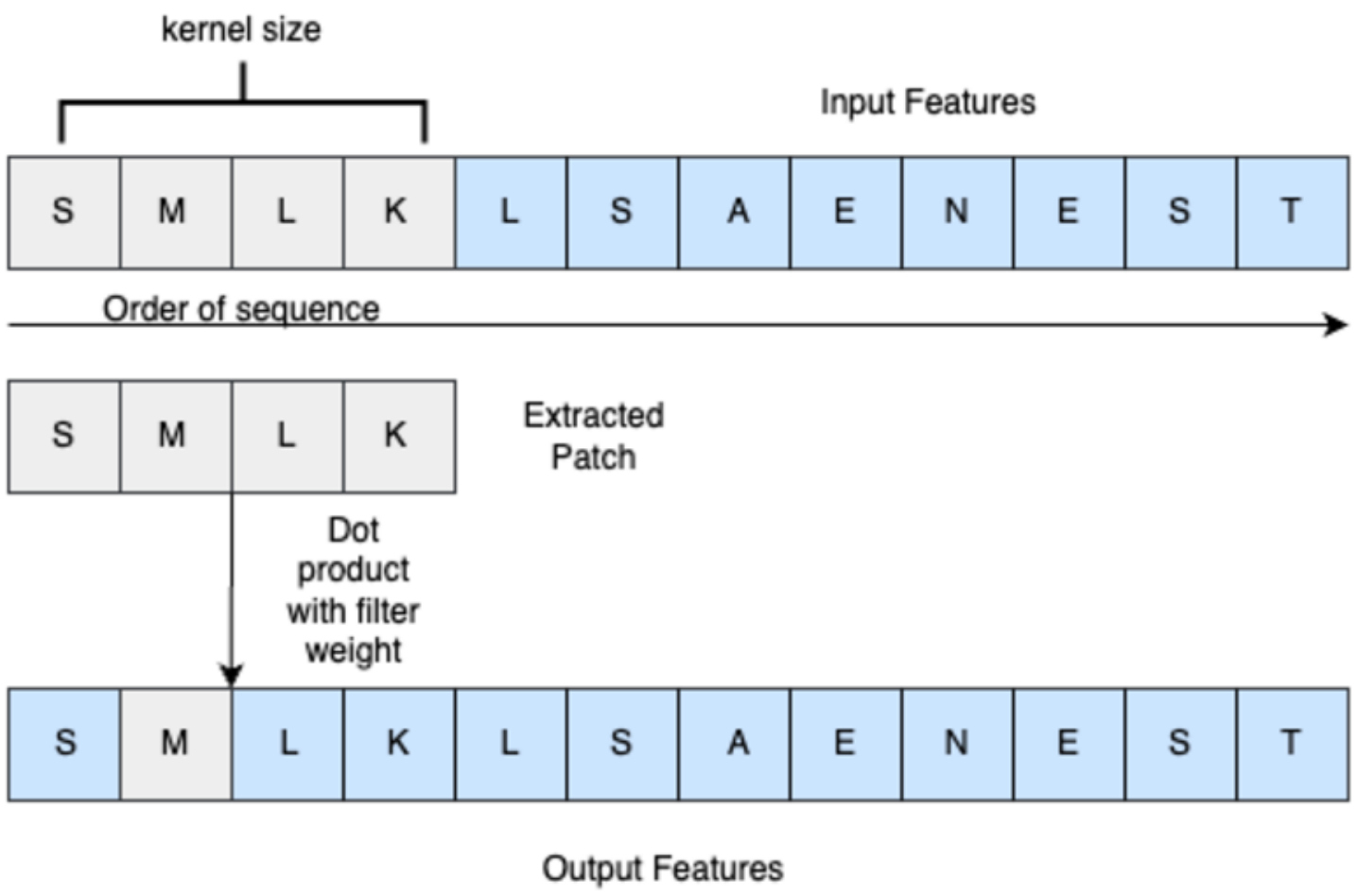

- **Architecture**: A multichannel architecture with three convolutional layers is used for Convolutional neural network implementation. There are 3 channels. The filter size is 128 and the kernel size is 12 in the first channel. In the second channel filter size is 64 and the kernel size is 6 while in the third channel filter size and kernel size are 32 and 3 respectively. These channels act as varying "n-grams" in NLP, enabling the model to learn features of different lengths within the sequence, from short motifs to longer patterns.
- **Convolution Operation**: Each filters slide over the input sequence, performing element-wise multiplications and summations to produce feature maps. The filters capture local patterns and dependencies within the sequence of kernel size like n-gram. Each filter learns different features, allowing the model to capture diverse aspects of the input sequence. The filters with different widths help to capture patterns of varying lengths. The **convolution unit** functions similarly to a Position-Specific Scoring matrix (PSSM), capturing patterns that are specific to each protein family. It learns these patterns through training, without predefined alignment.
- **Non-Linearity with ReLU**: The Rectified Linear Unit (ReLU) activation function is applied after each convolution operation, introducing non-linearity to the model. ReLU sets all negative values to zero, which allows the network to model more complex relationships.
- **Dropout Regularization**: A dropout layer with a rate of 0.2 is applied to prevent overfitting by randomly "dropping" 20% of neurons during each training step. This means that the output of those neurons is temporarily ignored or set to zero. This enhances the model's generalizability.
- **Pooling and Dimensionality Reduction**: Maxpooling1D layer pool size is kept at 2. It selects the maximum value within a fixed window size, reducing the dimensionality of the feature maps while preserving the most salient information.
- **Flattening and Dense Layers**: The pooled feature maps from all three channels are flattened and concatenated into a single one-dimensional vector, which is then passed through dense layers to combine the extracted features for classification. The final output layer with softmax activation produces probabilities for each protein class. The model is trained using categorical cross entropy and is compiled using Adam optimizer.

An outline of the proposed implementation of a convolutional neural network is drawn in (Fig. 6). Hyperparameters used in this study mentioned in Table 3.

**Fig. 6 Multi-channel CNN implementation outline**

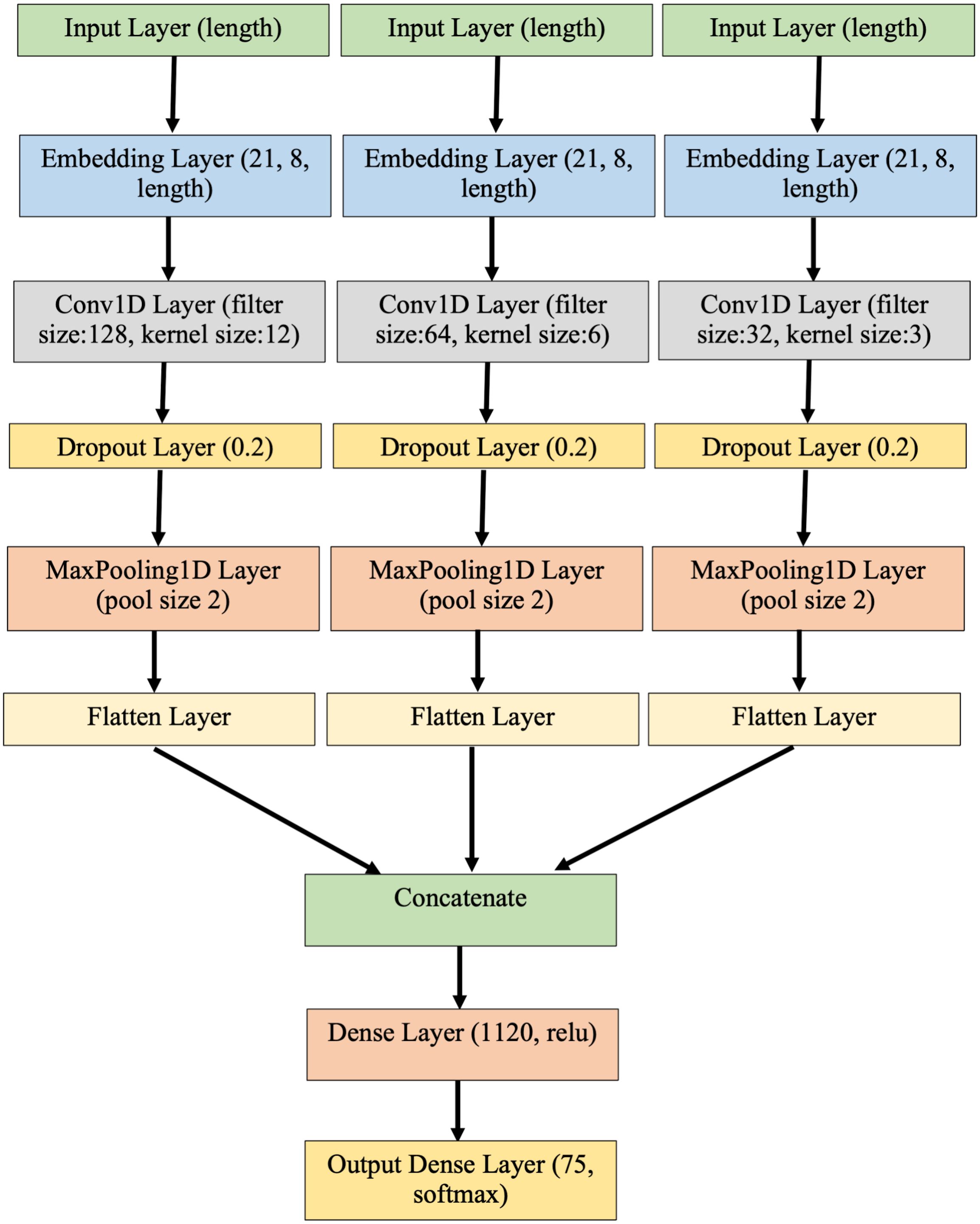

**Table 3 Hyperparameters for deep learning models**

| Hyperparameter | Values for LSTM | Values for CNN 1D |
|---|---|---|
| No. of Epochs | 30 | 20 |
| Learning rate | 0.001 | 0.001 |
| Batch size | 32 | 128 |
| Optimizer | Adam | Adam |
| Max length | 100, 256, 512 | 100, 256, 512 |
| Loss function | Categorical Cross entropy loss | Categorical Cross entropy loss |

## 3.5 Transformer models

Transformer models, based on attention mechanisms, have been highly effective in natural language processing and are increasingly being adapted for biological sequence analysis. In this study, we employed three transformer-based models: **BertForSequenceClassification**, **DistilBERT**, and **ProtBERT**. BertForSequenceClassification and DistilBERT are general-purpose transformer models adapted for sequence classification, while ProtBERT is a protein-specific transformer trained on extensive protein datasets, making it particularly suited for biological sequence analysis. ProtBERT leverages the self-supervised learning paradigm to train on large protein sequence corpora, enabling it to decode complex biological information embedded in protein sequences.

### 3.5.1 BertForSequenceClassification

#### Implementation steps

a. **Data pre-processing:** The pre-processed PDB protein dataset, consisting of protein sequences and their labels, was split into training, validation, and testing sets (80:10:10). Protein sequences were tokenized using the BERT tokenizer, which converts each sequence into tokens. These tokens were then transformed into input features (token IDs, attention masks, and segment IDs) that help BERT understand relationships within the sequences.
b. **Model initialization:** The model was initialized with pre-trained weights from BERT, capturing knowledge from a large corpus of text data to enhance performance on the protein classification task.
c. **Training loop:** Hyperparameters for training, such as learning rate (2e-5), batch size (4), and number of epochs (30), were set. During each epoch, the model learned from the protein sequences and their corresponding labels in batches. For each batch, a forward pass was performed, obtaining predicted logits. The cross-entropy loss between the predicted logits and true labels was computed to measure performance. A backward pass computed gradients of the loss with respect to the model's parameters. The optimizer, AdamW, updated the model's parameters based on the gradients to minimize the loss.
d. **Validation:** After training, the model's performance was evaluated on the validation set containing unseen protein sequences. A forward pass was performed to obtain predicted logits, which were compared to true labels to assess accuracy, precision, recall, and F1-score.

e. **Fine-tuning and optimization:** If performance was suboptimal, hyperparameters were fine-tuned or different optimization techniques were experimented with. The training loop was repeated with updated settings until the desired performance was achieved.
f. **Evaluation and save model:** Once trained and optimized, the model's weights and architecture were saved for future use. Evaluation was performed on testing data to determine the model's performance on unseen data.

### 3.5.2 DistilBERT

## Implementation steps

a. **Data preparation and neural network for fine-tuning:** The DistilBert tokenizer fast was used to prepare the protein sequences. A neural network was built with the DistilBERT model, followed by a dropout layer and a linear output layer to produce final predictions. The protein sequence data was passed through the DistilBERT model, and the outputs were compared to the true labels to compute model accuracy. An instance of this network, referred to as the model, was created for training and subsequent inference.
b. **Loss Function and optimizer:** The cross-entropy loss was calculated to assess performance. An optimizer, AdamW, updated the weights of the neural network to enhance accuracy and efficiency.
c. **Model training:** The model was trained for 30 epochs with batch training. During training, predictions were compared to the true labels to calculate loss, and the optimizer updated the model's weights. Validation was performed on unseen data to assess generalization, and performance metrics (e.g., accuracy) were evaluated.
d. **Model validation:** During validation, unseen validation data was passed to the model to assess performance and generalization to new data.
e. **Model evaluation:** The model was evaluated using a separate testing dataset to determine performance on entirely new and unseen data, crucial for understanding its real-world applicability.

### 3.5.3 ProtBert

The implementation steps for ProtBERT are similar to those for BertForSequenceClassification. However, ProtBERT is specifically pre-trained on protein sequences, making it more effective in capturing the nuances of biological data. The same procedure of data pre-processing, model training, and evaluation was followed.

The flowchart in Fig. 7 shows the experimental setup for three BERT approaches (BertForSequenceClassification, DistilBert and ProtBert).

**Fig. 7 Flow chart for BERT experimental setup**

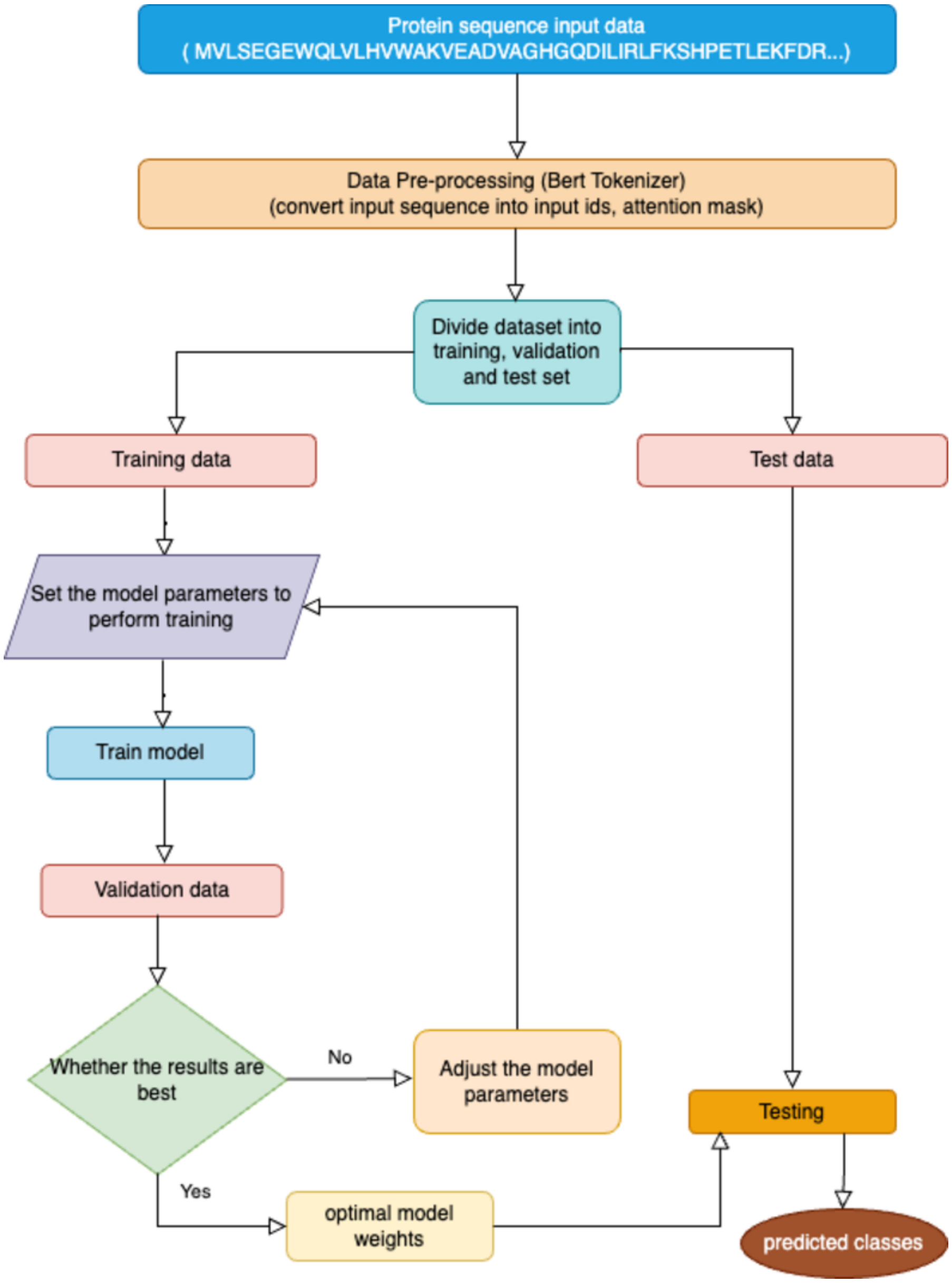

Parameters used for training Bert models are mentioned in Table 4. All three models were trained over 30 epochs, with performance metrics recorded after every 10 epochs. Due to computational constraints, the training was conducted in three separate intervals. The mean and standard deviation of the accuracy, f1 score and loss over these intervals provide a comprehensive overview of the model's performance.

**Table 4 Parameters for Bert models**

| Parameter | Values |
|---|---|
| No. of Epochs | 30 |
| Learning rate | 2e-5 |
| Batch size | 4 |
| Optimizer | AdamW |
| Max length | 281 |
| Loss function | Cross entropy loss |

# 4. Results

## 4.1 Machine learning

Training results of the best model are included in Table 5. The training results of various models demonstrate diverse performance metrics. The **4-gram Voting Soft** model achieved the highest accuracy (0.726 ± 0.006) and weighted F1 score (0.728 ± 0.003), indicating robust overall performance. The **4-gram Stacking** model also performed well, though slightly below Voting Soft in accuracy (0.716 ± 0.006) and weighted F1 score (0.718 ± 0.005). The **2-gram XGB** and **4-gram MLP** models showed strong performance, particularly in accuracy and weighted F1 scores. Conversely, models like **4-gram NB** and **4-gram DTC** exhibited lower performance metrics. Overall, ensemble methods like Voting Soft and Stacking outperformed individual models, showcasing their effectiveness in improving predictive performance across different metrics for imbalanced datasets.

When applying **ECOD-based splitting**, model performance generally decreased, suggesting a more challenging dataset split. The **4-gram Voting Soft model** remained the best performer but showed a noticeable drop, achieving **0.666 ± 0.08 accuracy** and **0.648 ± 0.003 weighted F1-score**. Similarly, the **4-gram Stacking model** saw a reduction to **0.638 ± 0.004 accuracy** and **0.622 ± 0.005 weighted F1-score**. The **2-gram XGB (0.638 ± 0.001 accuracy, 0.614 ± 0.001 weighted F1-score)** and **4-gram MLP (0.621 ± 0.001 accuracy, 0.605 ± 0.001 weighted F1-score)** models also experienced performance degradation. Notably, classifiers such as **4-gram NB (0.457 ± 0.16 accuracy, 0.441 ± 0.12 weighted F1-score)** and **3-gram KNN (0.582 ± 0.156 accuracy, 0.561 ± 0.124 weighted F1-score)** exhibited a significant drop in predictive ability. This trend underscores the increased difficulty posed by ECOD splitting, as well as the robustness of ensemble methods in maintaining relatively higher performance compared to individual models.

**Table 5 Result of machine learning approach training**

| **Models** | **Accuracy (mean±std)** | **f1-score Weighted (mean±std)** | **Accuracy ECOD (mean±std)** | **f1-score Weighted ECOD(mean±std )** |
|---|---|---|---|---|
| **3-gram KNN** | 0.652 ± 0.003 | 0.687 ± 0.018 | 0.582 ± 0.156 | 0.561 ± 0.124 |
| **4-gram Multinomial NB** | 0.617 ± 0.002 | 0.635 ± 0.001 | 0.457± 0.16 | 0.441 ± 0.12 |
| **4-gram LR** | 0.675 ± 0.003 | 0.684 ± 0.003 | 0.595 ± 0.003 | 0.574 ± 0.003 |
| **4-gram MLP** | 0.691 ± 0.001 | 0.685 ± 0.001 | 0.621 ± 0.001 | 0.605 ± 0.001 |
| **4-gram DTC** | 0.621 ± 0.001 | 0.634 ± 0.002 | 0.554 ± 0.008 | 0.534 ± 0.006 |
| **4-gram RFC** | 0.648 ± 0.001 | 0.648 ± 0.001 | 0.628 ± 0.001 | 0.603 ± 0.002 |
| **2-gram XGB** | 0.692 ± 0.002 | 0.685 ± 0.002 | 0.638 ± 0.001 | 0.614 ± 0.001 |
| **4-gram Voting Soft** | 0.726 ± 0.006 | 0.728 ± 0.003 | 0.666 ± 0.08 | 0.648 ± 0.003 |
| **4-gram Stacking** | 0.716 ± 0.006 | 0.718 ± 0.005 | 0.638 ± 0.004 | 0.622 ± 0.005 |

Table 6 presents a detailed comparison of different machine learning models' performance on the test data, analysed across different n-gram ranges (uni-gram, bi-gram, tri-gram, and 4-gram). The evaluation metrics include accuracy, macro F1 score, and weighted F1 score, both **with and without ECOD-based splitting**, providing insights into how each model performs under varying textual representations.

**Table 6 Result of machine learning approach (Test data)**

| **Models** | **Accuracy (mean±std)** | **f1-score Weighted (mean±std)** | **Accuracy ECOD (mean±std)** | **f1-score Weighted ECOD(mean±std )** |
|---|---|---|---|---|
| **KNN** | 0.71 ± 0.02 | 0.73± 0.01 | 0.54 ± 0.01 | 0.51 ± 0.01 |
| **Multinomial NB** | 0.38 ± 0.18 | 0.39 ± 0.19 | 0.30± 0.11 | 0.29 ± 0.11 |
| **LR** | 0.45 ± 0.25 | 0.45 ± 0.26 | 0.36 ± 0.17 | 0.34 ± 0.17 |
| **MLP** | 0.67 ± 0.06 | 0.66 ± 0.07 | 0.49 ± 0.06 | 0.47 ± 0.06 |
| **DTC** | 0.66 ± 0.01 | 0.66 ± 0.01 | 0.46 ± 0.04 | 0.45 ± 0.04 |
| **RFC** | 0.71 ± 0.00 | 0.71 ± 0.02 | 0.54 ± 0.02 | 0.52 ± 0.02 |
| | | | | |
| **XGB** | 0.71 ± 0.02 | 0.71 ± 0.01 | 0.56 ± 0.02 | 0.54 ± 0.02 |
| **Voting Soft** | 0.74 ± 0.01 | 0.74 ± 0.01 | 0.59 ± 0.02 | 0.56 ± 0.03 |
| **Stacking** | 0.74 ± 0.01 | 0.74 ± 0.00 | 0.59 ± 0.03 | 0.55 ± 0.03 |

Among all models, Voting Soft and Stacking ensembles demonstrated the best performance, achieving an accuracy of 0.74 ± 0.01 and a weighted F1-score of 0.59 ± 0.02 (without ECOD). These ensemble techniques effectively improved classification accuracy, indicating their ability to generalize well across different text representations. Similarly, the XGB, RFC, and KNN models performed competitively, with accuracy values of 0.71 ± 0.02 and weighted F1-scores ranging from 0.54 to 0.56, highlighting their potential for handling complex textual data.

In contrast, models such as Multinomial NB (0.38 ± 0.18 accuracy, 0.39 ± 0.19 weighted F1-score) and Logistic Regression (0.45 ± 0.25 accuracy, 0.45 ± 0.26 weighted F1-score) exhibited significantly lower performance, indicating their limitations when dealing with more complex n-gram structures. Similarly, DTC (0.66 ± 0.01 accuracy, 0.46 ± 0.04 weighted F1-score) and MLP (0.67 ± 0.06 accuracy, 0.49 ± 0.06 weighted F1-score) models performed moderately, showing some effectiveness but falling short compared to ensemble-based approaches.

## Impact of ECOD-Based Splitting on Test Performance

With ECOD-based splitting, a general decrease in performance was observed across all models. Voting Soft and Stacking ensembles, while still leading, saw reductions in accuracy (0.56 ± 0.03 weighted F1-score for Voting Soft and 0.55 ± 0.03 for Stacking). Similarly, the XGB model dropped to 0.54 ± 0.02 weighted F1-score, and RFC declined to 0.52 ± 0.02 weighted F1-score. The decline suggests that ECOD-based splitting creates a more challenging data distribution, affecting generalizability.

The Multinomial NB and Logistic Regression models were particularly affected, with their accuracy falling to 0.30 ± 0.11 and 0.36 ± 0.17, respectively. This drop highlights their sensitivity to distribution shifts introduced by ECOD. Conversely, models like KNN (0.54 ± 0.01 weighted F1-score) and XGB (0.54 ± 0.02 weighted F1-score) showed relatively better stability, suggesting their capability to adapt under different test conditions.

Fig. 8 presents a visual comparison of various machine learning models' performance across different n-gram ranges (uni-gram, bi-gram, tri-gram, and 4-gram).

**Fig. 8 Comparison of models' accuracy and f1 score**

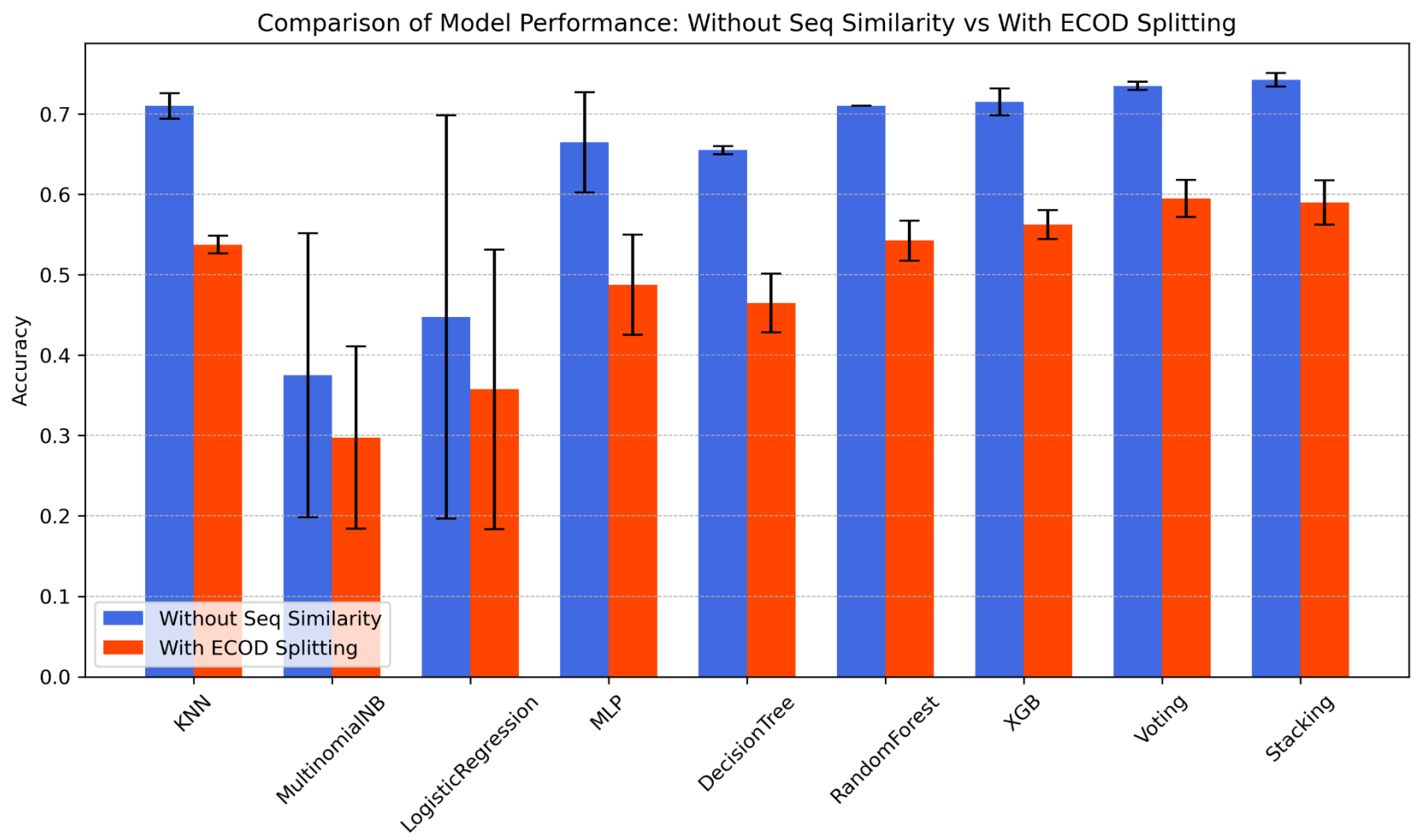

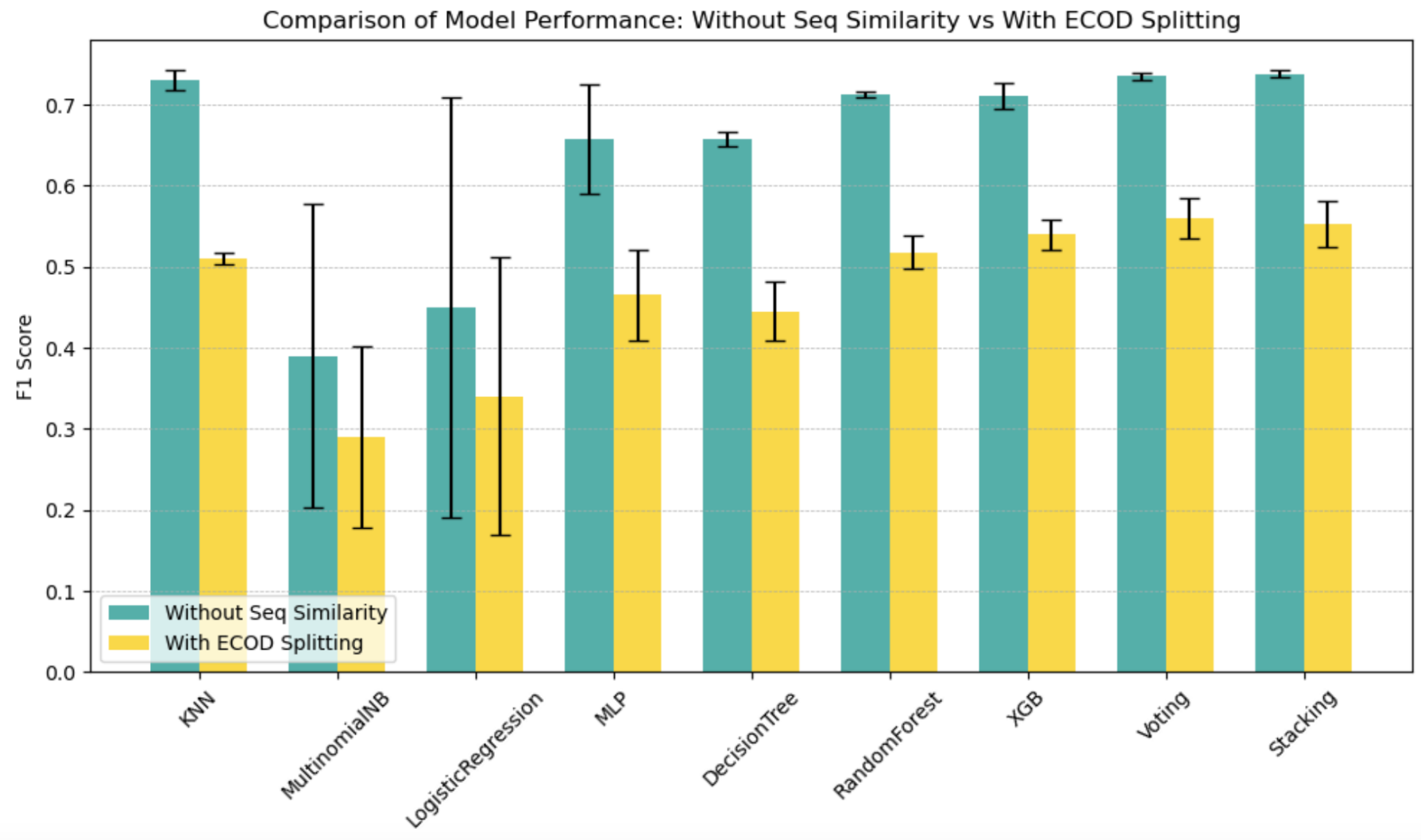

## 4.2 Deep learning

6 experiments are carried out for three different protein sequence lengths with and without class weight for CNN and 3 experiments for LSTM. The LSTM model shows moderate training and validation accuracy, indicating a reasonable but not optimal fit to both the training and validation data. The loss values indicate a moderate level of error in predictions. The F1 scores, particularly the macro F1 score, suggest that while the model performs reasonably well on more frequent classes, its performance on less frequent classes is less reliable and quite variable (Table 7 and Fig. 9).

**Table 7 Training result of LSTM and CNN models**

| Performance Metrics | Result (LSTM 100 Max length) | Result (CNN 256 Max length) | Result ECOD (LSTM 100 Max length) | Result ECOD (CNN 256 Max length) |
|---|---|---|---|---|
| **Training Accuracy (mean ± std)** | 0.442 ± 0.098 | 0.739 ± 0.154 | 0.422 ± 0.098 | 0.703 ± 0.170 |
| **Training Loss (mean ± std)** | 2.104 ± 0.397 | 0.938 ± 0.619 | 2.104 ± 0.397 | 0.989 ± 0.656 |
| **Validation Accuracy (mean ± std)** | 0.431 ± 0.082 | 0.640 ± 0.037 | 0.410 ± 0.082 | 0.610 ± 0.045 |
| **Validation Loss (mean ± std)** | 2.182 ± 0.300 | 1.497 ± 0.083 | 2.182 ± 0.300 | 1.245 ± 0.150 |
| **Validation F1 Score (weighted) (mean ± std)** | 0.424 ± 0.197 | 0.632 ± 0.042 | 0.402 ± 0.197 | 0.600 ± 0.052 |
| **Validation F1 Score (macro) (mean ± std)** | 0.310 ± 0.202 | 0.510 ± 0.068 | 0.310 ± 0.202 | 0.490 ± 0.081 |

**Fig. 9 Accuracy and loss for LSTM model for different sequence lengths (512, 256, 100)**

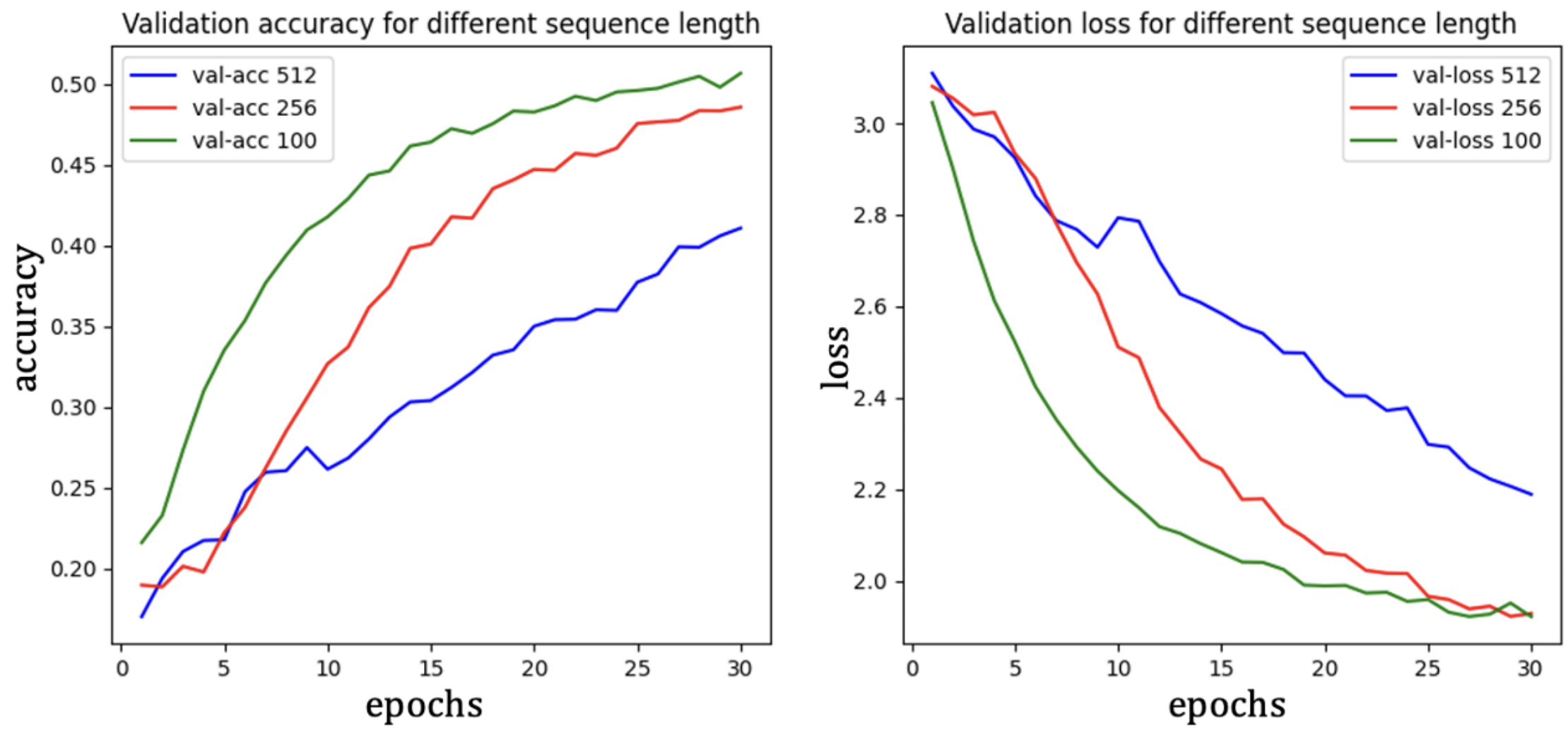

The CNN model shows good training accuracy, indicating effective learning, though with notable performance variability. The training loss reflects accurate predictions but inconsistent errors. Validation metrics indicate moderate generalization with more stable performance than training. F1 scores suggest the model handles frequent classes well, but performance varies across different classes (Table 7 and Fig. 10).

**Fig. 10 Accuracy and loss for CNN model for different sequence lengths (512, 256, 100)**

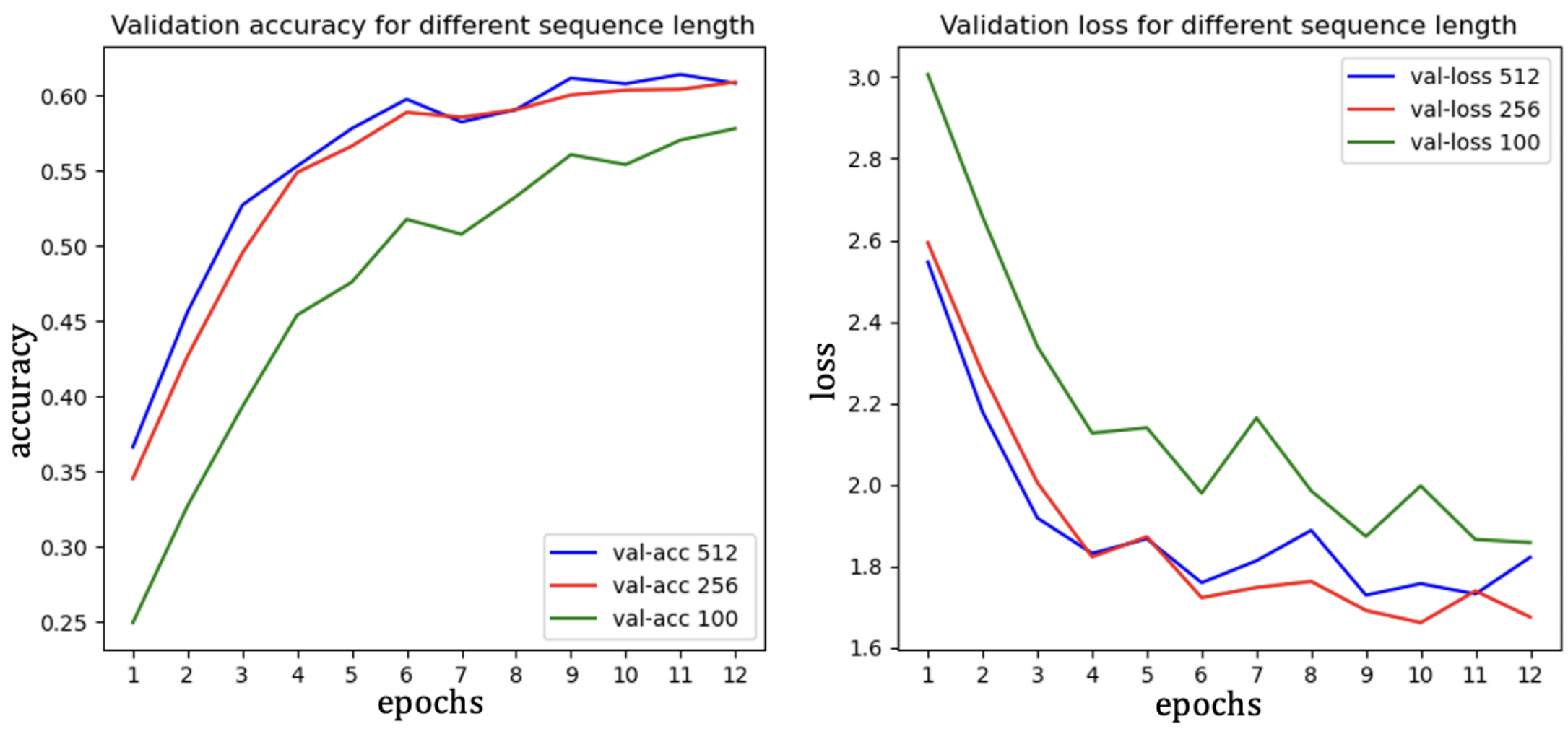

Fig. 10 illustrates that the CNN model's accuracy and loss trends are consistent for sequence lengths of 256 and 512, indicating stable model performance across these lengths. However, when the sequence length is reduced to 100, the model struggles to accurately classify protein sequences, resulting in higher loss values. This increased loss suggests a significant drop in model performance, highlighting that shorter sequence lengths are less effective for identifying the correct class in protein sequence classification tasks. The data suggests that longer sequences provide more information, leading to more reliable model predictions and better overall performance.

According to Table 8, The CNN model consistently outperformed the LSTM model across different sequence lengths, with the optimal performance observed at a sequence length of 256. At this length, the CNN achieved an accuracy of 67%, a test loss of 1.48, and a macro F1 score of 55%, making it the best performer among the configurations tested. When class weights were applied, the CNN's accuracy slightly decreased to 61%, but the macro F1 score improved marginally to 53%. In contrast, the LSTM model showed its best performance at a shorter sequence length of 100, where it achieved an accuracy of 51.0%, a macro F1 score of 48.0%, and a test loss of 1.92. However, the LSTM's performance deteriorated as the sequence length increased, with accuracy dropping to 41.0% and the macro F1 score falling sharply to 16.0% at a length of 512. When evaluating the models **with ECOD**, performance across all metrics declined. The best ECOD result for CNN was observed at 256 sequence length, where it achieved an accuracy of 52.0%, a macro F1 score of 42.0%, and a weighted F1 score of 48.0%. Similarly, the best ECOD result for LSTM was at 100 sequence length, with an accuracy of 43.0%, a macro F1 score of 27.0%, and a weighted F1 score of 39.0%.

**Table 8 Result of LSTM and CNN for three different sequence lengths on test data**

| **Metrics** | **100** | **256** | **512** | **ECOD (100)** | **ECOD (256)** | **ECOD (512)** |
|---|---|---|---|---|---|---|
| **LSTM Test Loss** | **1.92** | 1.93 | 2.20 | **1.92** | 1.92 | 2.30 |
| **LSTM Test Accuracy (%)** | **51.0** | 49.0 | 41.0 | **43.0** | 40.0 | 39.0 |
| **LSTM Test Macro F1 Score (%)** | **32.0** | 28.0 | 16.0 | **27.0** | 25.0 | 15.7 |
| **LSTM Test Weighted F1 Score (%)** | **48.0** | 46.0 | 36.0 | **39.0** | 37.0 | 34.0 |
| **CNN Test Loss** | 1.58 | **1.48** | 1.50 | 1.84 | **1.71** | 2.18 |
| **CNN Accuracy (%)** | 65.0 | **67.0** | 66.0 | 52.0 | **52.0** | 50.0 |
| **CNN Macro F1 Score (%)** | 50.0 | **55.0** | 55.0 | 37.0 | **42.0** | 41.0 |
| **CNN Weighted F1 Score (%)** | 64.0 | **66.0** | 66.0 | 43.0 | **48.0** | 47.0 |
| **CNN Test Loss with Class Weight** | 1.78 | 1.70 | 1.71 | 2.22 | 1.88 | 2.04 |
| **CNN Accuracy with Class Weight (%)** | 60.0 | 61.0 | 62.0 | 45.0 | 49.0 | 48.0 |
| **CNN Macro F1 Score with Class Weight (%)** | 52.0 | 53.0 | 53.0 | 33.0 | 40.0 | 39.0 |
| **CNN Weighted F1 Score with Class Weight (%)** | 61.0 | 62.0 | 63.0 | 41.0 | 46.0 | 46.0 |

It is clearly visible that the CNN model without class weight demonstrates superior performance in terms of accuracy and F1 scores, indicating better overall classification capability. The CNN model with class weight shows a slight decrease in performance, suggesting that while class weighting can help in addressing class imbalances, it may also introduce complexities that reduce overall effectiveness. The LSTM model underperforms compared to both CNN models, highlighting its limitations in identifying hidden patterns in longer sequences.

Fig. 11 presents a performance comparison of CNN and LSTM models across sequence lengths (100, 256, and 512) under two conditions: without sequence similarity filtering and with ECOD filtering. For CNN, accuracy remains stable without ECOD (blue line), peaking at 256 sequence length (0.67) and staying above 0.65, whereas with ECOD (green line), accuracy starts at 0.52 and declines slightly to 0.50, indicating classification challenges introduced by ECOD. In terms of weighted F1-score, CNN maintains a strong range of 0.64 to 0.66 without ECOD, but with ECOD, it starts lower (0.37), improves at 256 sequence length (0.42), and stabilizes at 0.41 at 512, suggesting that ECOD may balance predictions but at the cost of overall classification performance. For LSTM, accuracy declines consistently without ECOD (0.51 → 0.41) as sequence length increases, indicating reduced classification effectiveness for longer sequences. With ECOD, accuracy starts lower at 0.44 and further declines to 0.39, showing that ECOD filtering impacts LSTM's ability to generalize. Similarly, LSTM's weighted F1-score drops from 0.48 to 0.36 without ECOD as sequence length increases, and with ECOD, it remains lower (0.37 to 0.34), highlighting LSTM's struggle with longer sequences and ECOD-filtered data.

**Fig. 11 Comparison of deep learning approach**

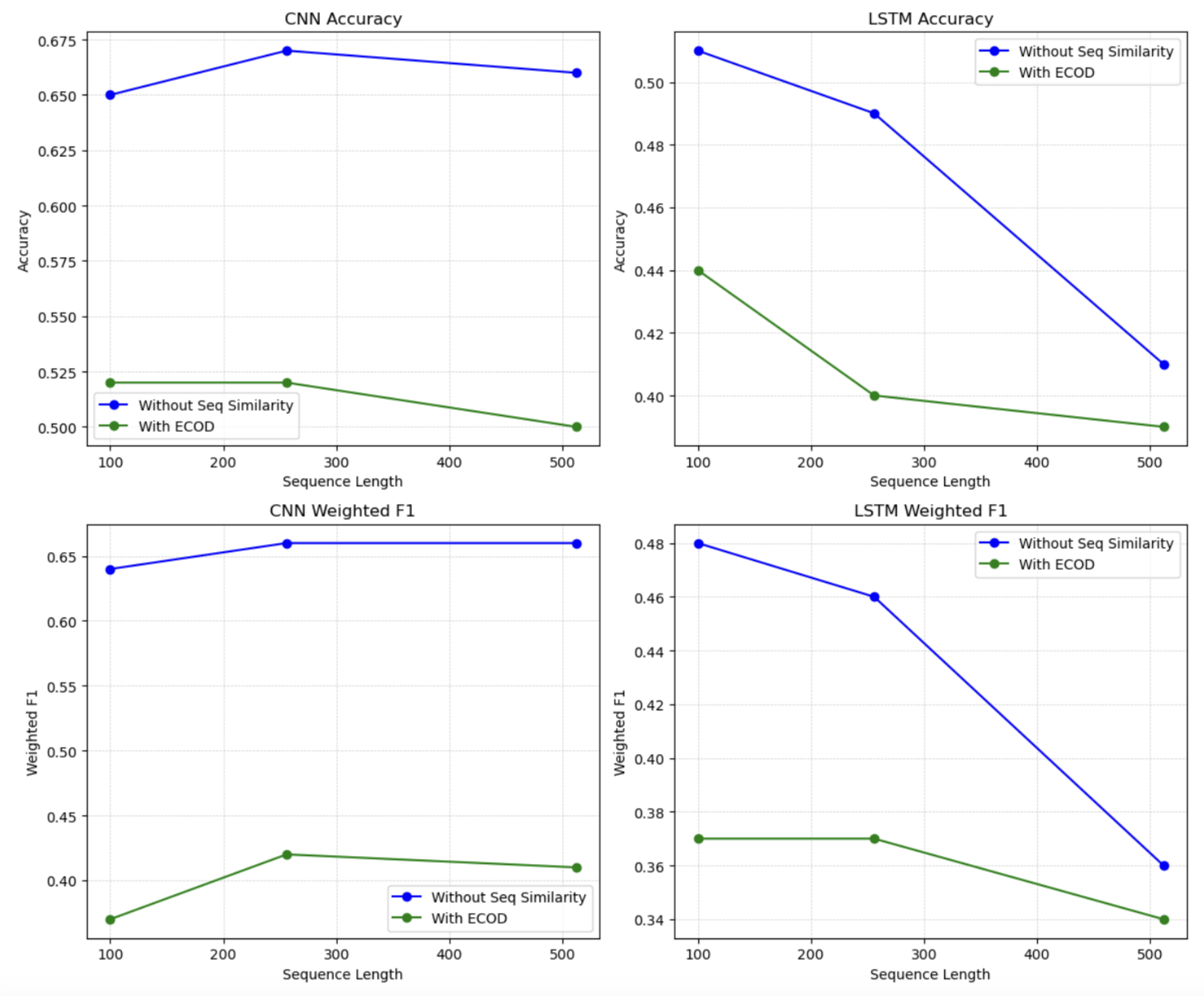

## 4.3 Transformer models

The training results for BERT models are summarized in Table 9. The Bert For Sequence Classification model demonstrated strong performance during training, with an average loss of 0.37 and an accuracy of 90.73%. However, it exhibited a noticeable drop in validation performance, where the validation loss increased to 1.7626, and accuracy and F1 weighted score dropped to approximately 72%. The DistilBERT model showed a slightly lower training performance, with a loss of 0.44 and an accuracy of 85%. Its validation accuracy was close to Bert For Sequence Classification's at 72%, with a validation loss of 1.4967, and an F1 score of 71.56%. The ProtBert model, while having the highest training loss at 0.79, demonstrated the best validation performance among the three, with a validation accuracy of 77.4% and an F1 weighted score of 75.42%, indicating its robustness on unseen data. These models achieved a good balance between training and validation performance up to 20 epochs, but beyond that, the improvement in training accuracy did not translate into better validation performance, suggesting the risk of overfitting.

**Table 9 Training result for BERT models**

| Model | Train Loss | Train Acc | Val Loss | Val Acc | Val f1 weighted |
|---|---|---|---|---|---|
| **Bert For Sequence Classification** | 0.367 ± 0.082 | 0.907 ± 0.025 | 1.763 ± 0.204 | 0.723 ± 0.009 | 0.720 ± 0.010 |
| **DistilBERT** | 0.443 ± 0.189 | 0.850 ± 0.049 | 1.497 ± 0.253 | 0.720 ± 0.007 | 0.716 ±0.009 |
| **ProtBert** | 0.790 ± 0.277 | 0.824 ± 0.087 | 1.415 ± 0.060 | 0.774 ± 0.083 | 0.754 ± 0.083 |

Table 10 presents the training results for various BERT-based models using ECOD-Based Splitting. The models tested include BERT for Sequence Classification, DistilBERT, and ProtBERT. For the BERT model, the training loss was 1.056 ± 0.675, with a training accuracy of 0.720 ± 0.165. The validation loss and accuracy were 1.583 ± 0.423 and 0.639 ± 0.139, respectively, while the validation F1 weighted score was 0.635 ± 0.137. DistilBERT exhibited a higher training loss of 1.320 ± 0.662, with a lower training accuracy of 0.633 ± 0.145. Its validation loss was 1.493 ± 0.536, validation accuracy 0.620 ± 0.125, and validation F1 weighted score 0.605 ± 0.138. ProtBERT had a training loss of 1.373 ± 0.424 and a training accuracy of 0.654 ± 0.126. Its validation loss was 1.457 ± 0.290, with a validation accuracy of 0.639 ± 0.107 and a validation F1 weighted score of 0.617 ± 0.106.

**Table 10 Training result for BERT models with ECOD-Based Splitting**

| Model | Train Loss | Train Acc | Val Loss | Val Acc | Val f1 weighted |
|---|---|---|---|---|---|
| **Bert For Sequence Classification** | 1.056 ± 0.675 | 0.720 ± 0.165 | 1.583 ± 0.423 | 0.639 ± 0.139 | 0.635 ± 0.137 |
| **DistilBERT** | 1.320 ± 0.662 | 0.633 ± 0.145 | 1.493 ± 0.536 | 0.620 ± 0.125 | 0.605 ±0.138 |
| **ProtBert** | 1.373 ± 0.424 | 0.654 ± 0.126 | 1.457 ± 0.290 | 0.639 ± 0.107 | 0.617 ± 0.106 |

Table 11 presents the performance of three different BERT-based transformer models on the test dataset, highlighting key metrics such as test loss, accuracy, and F1 scores. All three models show competitive accuracy, with ProtBert achieving the highest at 77%. BertForSequenceClassification and DistilBERT both have an accuracy of 73%. ProtBert has the lowest test loss (1.37), indicating better performance in terms of prediction error. DistilBERT shows a lower test loss (1.48) compared to BertForSequenceClassification (1.78). ProtBert again leads with a weighted F1 score of 76%, indicating it performs well across classes considering the class distribution. BertForSequenceClassification and DistilBERT had similar performances with slight variations. BertForSequenceClassification and DistilBERT have similar weighted F1 scores, 73% and 72%, respectively. All models have the same macro

F1 score of 61%, reflecting their balanced performance across all classes, irrespective of class distribution.

When incorporating ECOD filtering, a decline in performance is observed across all models. BertForSequenceClassification's accuracy drops from 73% to 60%, with a weighted F1 score decreasing to 56% and macro F1 to 52%. Similarly, DistilBERT's accuracy reduces to 59%, and its weighted and macro F1 scores decline to 53% and 49%, respectively. ProtBert remains the best performer under ECOD filtering, though its accuracy drops to 61%, weighted F1 to 56%, and macro F1 to 53%. These results indicate that while ECOD affects all models, ProtBert retains relatively better performance compared to the others.

**Table 11 Test result for BERT models**

| Transformer Model | Test Loss | Accuracy | F1 score weighted | F1 score macro | Test Loss (ECOD) | Accuracy (ECOD) | F1 score weighted (ECOD) | F1 score macro (ECOD) |
|---|---|---|---|---|---|---|---|---|
| **BertForSequenceClassification** | 1.78 | 0.73 | 0.73 | 0.61 | 1.63 | 0.60 | 0.56 | 0.52 |
| **DistilBERT** | 1.48 | 0.73 | 0.72 | 0.61 | 1.52 | 0.59 | 0.53 | 0.49 |
| **ProtBert** | **1.37** | **0.77** | **0.76** | **0.61** | **1.53** | **0.61** | **0.56** | **0.53** |

Fig. 12 presents a comparative evaluation of multiple models in terms of accuracy, weighted F1-score, and macro F1-score, both with and without ECOD filtering. The top plot shows accuracy, where transformer models (BertForSequenceClassification, DistilBERT, and ProtBert) outperform traditional models, with ProtBert achieving the highest accuracy (77%). However, ECOD filtering leads to a performance decline across all models, with the most notable drop observed in deep learning models. The middle plot compares the weighted F1-score, showing that ProtBert maintains the highest score (76%) without ECOD, while all models experience a decline under ECOD filtering. The bottom plot illustrates the macro F1-score, where transformer models maintain the best-balanced performance across classes. Overall, the results indicate that while ECOD filtering affects performance, ProtBert remains the most robust model.

**Fig. 12 Comparison of All models**

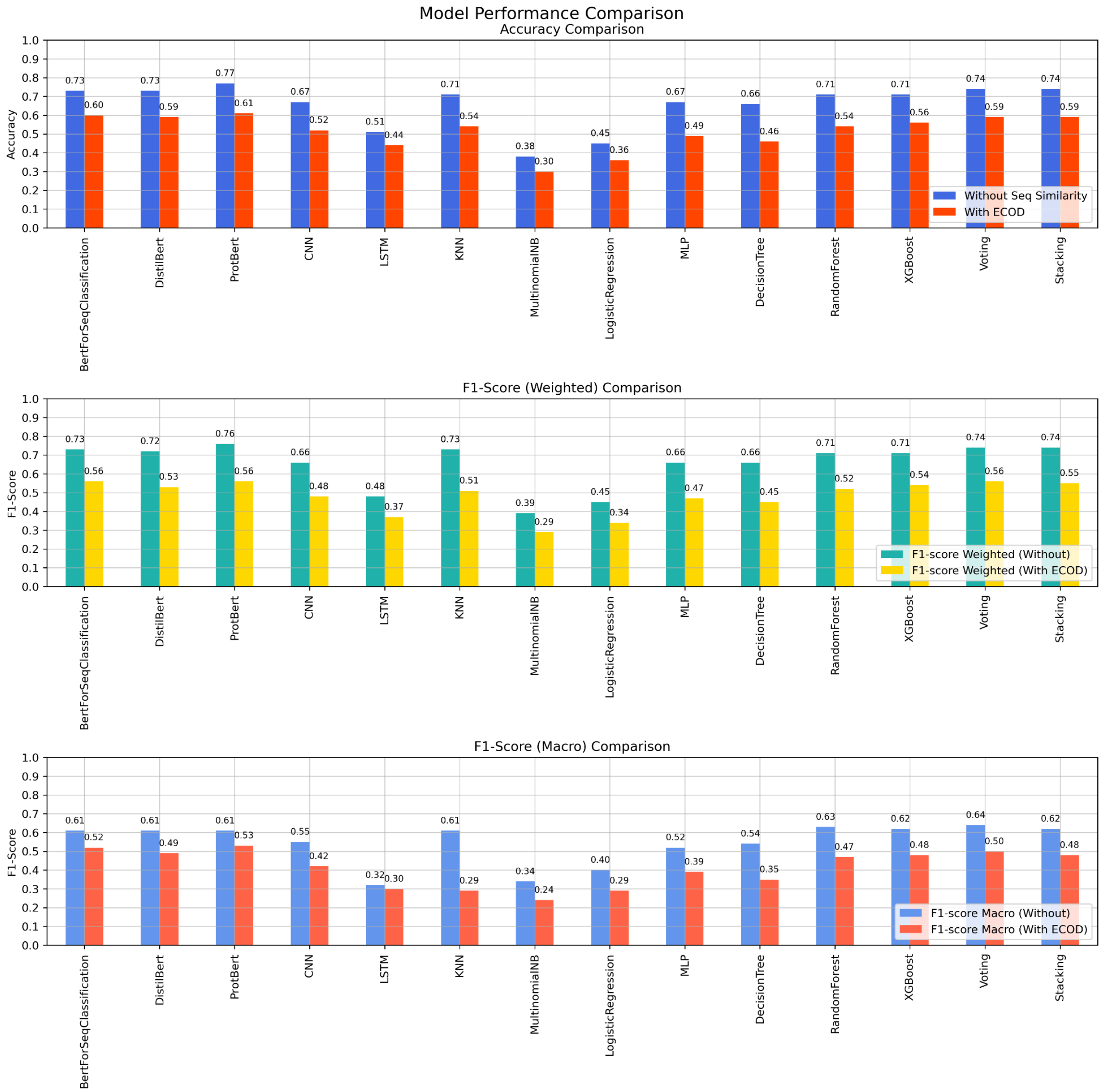

## 4.4 Error analysis

In general, proteins can be different types of enzymes, signalling proteins, structural proteins, and a variety of other options. Since many proteins are designed to bind in the same locations as one another, they frequently exhibit extremely similar properties. A Hydrolase enzyme and a Hydrolase inhibitor protein, for instance, will have similar structures since they focus on the same regions. Fig. 13 shows some examples of error analysis.

**Fig. 13 Error analysis**

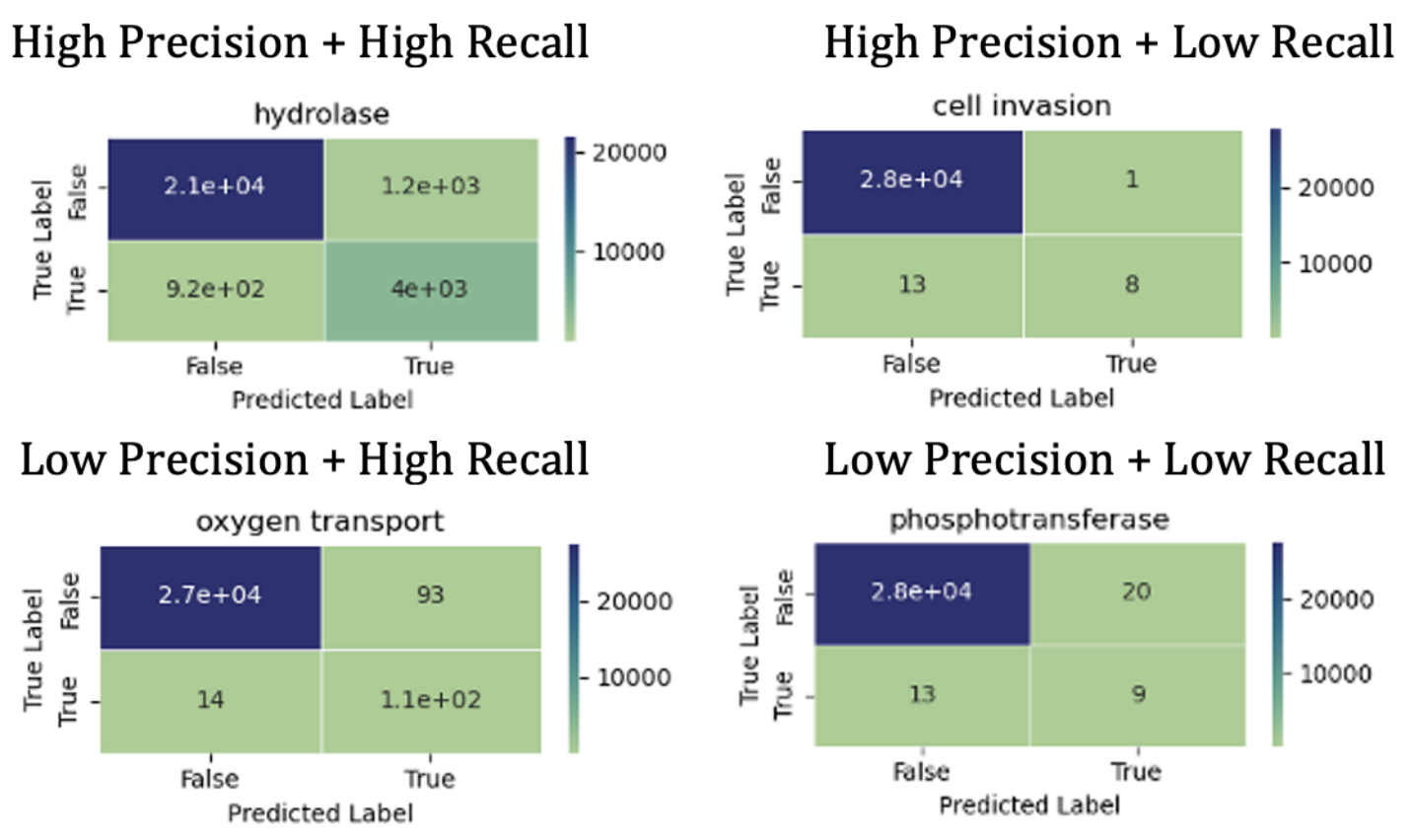

**High precision and high recall** that means model is able to identify them because these classes either have enough sample or their structure are not similar to others for instance, Allergan, apoptosis, immune system, isomerase, and hydrolase.

Classes have **high precision and low recall** that means model does not identify them correctly, For example, RNA binding proteins, DNA binding proteins, and transcription proteins, all share characteristics with gene regulator proteins and Cell invasion, Cell adhesion having similarity with cell cycle that make model difficult to identify them.

Classes are showing **low precision and high recall** like electron transport and oxygen storage, model is able to well detect these classes but also includes observations from other classes as well, for instance oxygen storage might be make model misleading for oxygen transport.

Classes e.g. Phosphotransferase, Transcription inhibitor have **low precision and low recall** which means model is not able to identify correct classes and model is not doing well on the entire test dataset to find correct classes. On the other side some classes like, Ribosome having low precision and low recall because of similarity with ribosomal protein (fundamental building blocks for ribosome) that means the model is not able to detect the correct class.

Error analysis reveals several areas where the model encounters challenges in distinguishing between closely related protein classes, primarily due to overlapping structural or functional properties. Below, we provide detailed insights into specific misclassification patterns observed:

1. **Overlap in Binding Sites and Functional Similarities**: Misclassifications frequently occur among RNA-binding proteins, DNA-binding proteins, and transcription-related proteins. These classes share functional roles in gene regulation, with structural similarities that often lead the model to group them together. This results in high precision but lower recall, as the model sometimes fails to recognize subtle distinctions between these classes. For example, transcription factors and RNA-binding proteins may exhibit similar motifs, causing occasional classification errors.
2. **Similarity in Protein Roles**: The model struggles to accurately classify proteins involved in cell adhesion, cell invasion, and cell cycle processes, as these classes have overlapping functions in cellular mechanisms. This overlap leads to moderate confusion across these classes, where the model's predictions reflect shared structural or sequence motifs. We observed that certain sequence characteristics common to these functions contribute to the model's difficulty in discerning these classes individually.

3. **Structural Parallels in Storage and Transport Proteins**: Classes such as electron transport and oxygen storage exhibit low precision but high recall. For instance, the model often confuses oxygen storage proteins with oxygen transport proteins due to similar sequence patterns associated with binding and storage. These overlaps highlight the challenges of distinguishing between functionally related but structurally distinct proteins.
4. **Challenges with Sparse Classes**: Certain classes, including Phosphotransferase and Transcription inhibitors, show both low precision and low recall. This issue appears linked to limited sample diversity and structural similarities with other enzyme classes in the training dataset. Additionally, ribosome-related proteins are often misclassified as ribosomal proteins due to shared structural elements, reflecting the model's difficulty in separating structurally analogous yet functionally distinct classes.

# 5. Discussion

The study demonstrates that NLP techniques can significantly enhance protein sequence classification. The use of n-grams proved effective in improving classifier performance, and ensemble methods showcased their potential in handling imbalanced datasets. While CNN outperformed LSTM in handling longer sequences, transformer models, particularly ProtBert, demonstrated superior accuracy and F1 scores, albeit with higher computational requirements. To enhance model performance, we explored hyperparameter tuning, regularization, and class balancing techniques. Ensemble methods such as Voting and Stacking were particularly effective, as they aggregate multiple model predictions to improve robustness against class imbalance.

The findings of this study highlight the significant potential of natural language processing (NLP) techniques in protein sequence classification. Given the growing amount of biological data, efficient and automated methods are essential. The results demonstrate that various machine learning models, when applied to amino acid n-grams, can achieve noteworthy accuracy and F1 scores, underscoring the effectiveness of this approach.

The K-Nearest Neighbors (KNN) algorithm performed particularly well on tri-gram data, indicating its strength in capturing local sequence similarities. In contrast, models like Logistic Regression, MLP, and Random Forests showed their highest performance with 4-gram data, suggesting that these models benefit from a broader contextual understanding provided by longer n-grams.

The transformer models, particularly ProtBert, showed promise with competitive F1 scores, although they required significant computational resources. This emphasizes the importance of access to high-performance computing facilities for training such models efficiently. The challenges faced due to limited GPU access and session expiry constraints highlight a practical limitation in the current study, suggesting a need for more robust computational infrastructure for future research.

Despite these promising results, several challenges remain. The primary source of error across models was the imbalanced dataset, with some classes having significantly fewer samples. This imbalance likely hindered the models' ability to generalize well across all classes. Future work could explore advanced techniques for handling class imbalance, such as data augmentation or more sophisticated weighting schemes.

The similarity between sequences from different classes also posed a challenge, potentially confusing the models and reducing classification accuracy. Advanced sequence embedding techniques or incorporating additional biological context could help mitigate this issue.

Furthermore, the results indicate that while CNNs and LSTMs showed reasonable performance, transformer models like BERT variants provided more consistent results across different sequence lengths and configurations. This suggests that transformers may offer a more robust framework for protein sequence classification, benefiting from their ability to capture long-range dependencies and contextual information effectively. Transformer models, particularly ProtBERT, is a protein-specific transformer trained on extensive protein datasets, outperformed due to their ability to capture long-range dependencies in amino acid sequences, which are critical for identifying functional similarities across protein classes.

To contextualize our findings, we compared the results with those reported in recent studies employing similar techniques for protein classification Table 1. In comparison, this study employed a diverse range of models, including traditional machine learning algorithms, CNNs, LSTMs, and transformer models such as ProtBERT, achieving a highest accuracy of 77% with ProtBERT and 74% with a voting classifier. While some previous studies achieved higher accuracy on specific datasets or task-specific metrics, our results underscore the versatility of ProtBERT for general protein classification tasks. Additionally, our approach emphasizes the integration of ensemble methods to address class imbalance, contributing to robust performance across diverse protein classes.

In this study, we incorporated ECOD-based splitting for protein sequences to enhance the evaluation of models. ECOD (Evolutionary Classification of Domain Structures) is a classification system that organizes protein structures into distinct domains based on their evolutionary relationships, offering a more accurate and biologically relevant framework for sequence analysis. By applying ECOD splitting, we ensured that the training and validation datasets consisted of distinct protein domain families, reducing potential bias and enabling the models to generalize better across different structural contexts. This approach allowed us to better assess the performance of the models in handling diverse protein sequences, highlighting their ability to capture evolutionary and structural variations within the data.

In summary, this study demonstrates the feasibility and effectiveness of using NLP techniques for protein sequence classification. The insights gained here pave the way for further exploration and optimization of these methods, with the potential to significantly enhance our ability to analyze and interpret complex biological data. Our Voting and ProtBERT models demonstrated performance on par with recent benchmark methods in protein classification. Compared to traditional sequence alignment methods, our approach offers improved efficiency, particularly in handling large-scale datasets. Future research should focus on addressing the challenges of class imbalance and sequence similarity, as well as leveraging more advanced computational resources to fully realize the potential of these techniques.

# 6. Conclusion

Assigning class weights effectively managed the imbalanced dataset. The best machine learning approach was the Voting Soft classifier, while CNN outperformed LSTM in deep learning methods. Among transformer models, ProtBert achieved the highest performance, highlighting the potential of advanced NLP techniques in protein sequence classification.
As bioinformatics advances uncover novel proteins, the need for precise, effective, and automated protein sequence classification methods becomes increasingly important. Identifying the family or class of a protein sequence, which indicates its function, remains a fundamental challenge in protein sequence analysis. Most previous studies have worked on protein secondary structure and sub-cellular location prediction, which motivated this study to explore various feature extraction methods using natural language processing (NLP) techniques to classify protein sequences.

In this research, we addressed dataset imbalance by assigning class weights and incorporated ECOD-based splitting for protein sequences to enhance the evaluation of our models. ECOD (Evolutionary Classification of Domain Structures) allowed for a more biologically relevant dataset by ensuring distinct protein domain families were used for training and validation, minimizing potential bias and improving the models' ability to generalize across different protein structures. The dataset included 75 target protein classes, and machine learning experiments were conducted using amino acid ranges of 1-4 grams. Among the models tested, the K-Nearest Neighbors (KNN) algorithm excelled with tri-gram data, achieving 70.0% accuracy and a macro F1 score of 63.0%. Multinomial Naïve Bayes achieved 62.0% accuracy and a 55.0% F1 score with 4-gram data. Logistic regression performed best with 4-grams, attaining 71.0% accuracy and a 62.0% F1 score. The Multi-Layer Perceptron (MLP) model reached 73.0% accuracy and a 60.0% F1 score on 4-gram data. The decision tree model achieved 66.0% accuracy and a 55.0% F1 score. The random forest classifier attained 71.0% accuracy and a 63.0% F1 score on 4-gram data. The XGBoost classifier performed well on bi-gram data, with 73.0% accuracy and a 64.0% macro F1 score. The voting classifier achieved 74.0% accuracy and a 65.0% F1 score, while the stacking classifier achieved 75.0% accuracy and a 64.0% macro F1 score.

For convolutional neural network (CNN) and long short-term memory (LSTM) models, experiments were conducted using sequence lengths of 512, 256, and 100. The CNN model showed the best performance with a sequence length of 256, achieving 67.0% accuracy and a 55.0% macro F1 score. The LSTM model had lower accuracy and F1 scores, with the highest scores for a sequence length of 100.
Finally, three transformer models were used: BertForSequenceClassification, DistilBERT, and ProtBert. The results were similar across these models, with BertForSequenceClassification achieving a macro F1 score of 61.0% and a weighted F1 score of 73.0%, DistilBERT achieving 72.0%, and ProtBert achieving 76.0% weighted F1 score, although ProtBert required more computational resources. Training ProtBert on a Mac M1 MPS environment took approximately 30 hours for one epoch. Limited GPU access and session expiry issues prevented running the models on Google Colab Pro. Transfer learning on a Mac showed that each epoch for BertForSequenceClassification and DistilBERT required about 5 hours.

The main sources of model error were limited sample training data for some classes and sequence similarity across different classes. By including ECOD-based splitting, we ensured a more biologically meaningful training and validation process. Addressing these issues is crucial for further improving protein sequence classification accuracy.

# 7. Acknowledgements

I would like to express my sincere gratitude to my supervisor, Dr. Julie Weeds, for her invaluable guidance, support, and encouragement throughout the course of this research. Her expertise and insights have greatly contributed to the development of this work.

# Data Availability Statements

The dataset is publicly available at https://www.kaggle.com/datasets/shahir/protein-data-set . All source codes used in this study can be found at https://github.com/humaperveen/PortfolioProjects/tree/main/PythonProjects/NLP/ProteinSeqClassification

# Statements and declarations

## Funding

No funding was received for conducting this study.

## Ethical Approval

Not applicable.

## Conflict of Interest

The authors have declared that no competing interests exist.

# Abbreviations

1. **NLP** - Natural Language Processing
2. **KNN** - K-Nearest Neighbors
3. **Multinomial NB** - Multinomial Naive Bayes
4. **LR** - Logistic Regression
5. **MLP** - Multi-Layer Perceptron
6. **DTC** - Decision Tree Classifier
7. **RFC** - Random Forest Classifier
8. **XGB** - XGBoost Classifier
9. **CNN** - Convolutional Neural Network
10. **LSTM** - Long Short-Term Memory
11. **PDB** - Protein Data Bank
12. **RCSB** - Research Collaboratory for Structural Bioinformatics
13. **UniProt** - Universal Protein Resource
14. **PIR** - Protein Information Resource center
15. **V-ELM** – Voting-based Extreme Learning Machine
16. **VOP-ELM** – Voting-based Optimal Pruned Extreme Learning Machine
17. **ELM** - Efficient Extreme Learning Machine
18. **SLFNs** - Single-layer Feedforward Neural Networks
19. **SaE-ELM** - Self-adaptive Evolutionary Extreme Learning Machine
20. **ESM** - Evolutionary Scale Modeling
21. **TAPE-Transformer** - Tasks Assessing Protein Embeddings Transformer
22. **pLMs** - Protein-specific Language Models